\documentstyle[11pt,aaspp4]{article}

\begin{document}
\title{Stability of the Forward/Reverse Shock System Formed by the\\
Impact of a Relativistic Fireball on an Ambient Medium}
\author{Xiaohu Wang, Abraham Loeb}
\affil{Astronomy Department, Harvard University, 60 Garden
Street, Cambridge, MA 02138; xwang,aloeb@cfa.harvard.edu}
\author{and Eli Waxman}
\affil{Department of Condensed Matter Physics, Weizmann
Institute, Rehovot 76100, Israel; waxman@wicc.weizmann.ac.il}

\begin{abstract}

We analyze the stability of a relativistic double (forward/reverse) shock
system which forms when the fireball of a Gamma-Ray Burst (GRB) impacts on
the surrounding medium. We find this shock system to be stable to linear
global perturbations for either a uniform or a wind ($r^{-2}$) density
profile of the ambient medium. For the wind case, we calculate analytically
the frequencies of the normal modes which could modulate the early
short-term variability of GRB afterglows.  We find that perturbations in
the double shock system could induce oscillatory fluctuations in the
observed flux on short (down to seconds) time scales during the early phase
of an afterglow.

\end{abstract}


\section{Introduction}

Gamma-ray bursts (GRBs) and their afterglows are most naturally described
by the relativistic ``fireball'' model (see e.g., Paczy\'{n}ski \& Rhoads
1993; Katz 1994; M\'{e}sz\'{a}ros \& Rees 1993, 1997; Waxman 1997a,b; Sari,
Piran, \& Narayan 1998). In this model, a compact source releases a large
amount of energy over a short time and produces a fireball which expands
relativistically as a thin shell. When the shell encounters the circumburst
medium, two shocks are formed -- a forward shock which propagates into the
circumburst medium and accelerates it, and a reverse shock which propagates
into the relativistic shell and decelerates it (M\'{e}sz\'{a}ros \& Rees
1992; Katz 1994; M\'{e}sz\'{a}ros, Rees, \& Papathanassiou 1994; Sari \&
Piran 1995; Sari, Narayan, \& Piran 1996). Later on, after a significant
mass of circumburst medium is accumulated, the shell approaches a
self--similar behaviour, as originally described by Blandford \& McKee
(1976), where there is only one forward shock propagating into the
circumburst medium. The circumburst medium could be either the interstellar
medium (ISM) or a progenitor wind.

The stability of the Blandford--McKee (1976) solution has been demonstrated
recently by Gruzinov (2000).  Here, we analyze the stability of a
forward/reverse relativistic shock system. This double shock system exists
during an important phase in the evolution of GRBs and its stability has
observational consequences. In particular, oscillations or instabilities
could translate to specific patterns of temporal variability in the
lightcurves of GRB afterglows.

In our linear perturbation analysis we generalize the ``thin shell'' method
first introduced by Vishniac (1983) in the non--relativistic regime. This
method simplifies the equations describing the stability of a spherical
shock when the wavelength of the perturbation is much larger than the
thickness of the shocked shell. In our relativistic treatment we focus on
global perturbations for which the wavelength is much larger than the
thickness of the forward/reverse shock system.  We consider the regime of
GRB parameters where the reverse shock is relativistic (although in reality
it may also be non-relativistic).  In \S 2 we derive the perturbation
equations for the forward/reverse shock system. In \S 3 we show the
analytical results for the wind case and the numerical results for both the
wind and ISM cases. Finally, in \S 4 we summarize our main conclusions.

\section{Linear Perturbation Equations} 

The interaction between a relativistically expanding shell and the
circumburst medium results in a double shock system, as shown in Figure
1. The system includes four distinct regions: the circumburst medium
(region 1) and the shocked circumburst medium (region 2) are separated by
the forward shock (shock 1), while the shocked material in the shell
(region 3) and the unshocked material in the shell (region 4) are separated
by the reverse shock (shock 2). Region 2 and 3 are separated by a contact
discontinuity. Our analysis is done in the observer frame where the
circumburst medium is at rest. We use a spherical coordinate system whose
origin is located at the center of the explosion. The radii of the contact
discontinuity, shock 1 and shock 2 are denoted by $R_0$, $R_1$ and $R_2$
respectively. We refer to the combination of region 2 and 3 as the layer
whose stability we consider. Similarly to Vishniac (1983) we make the thin
shell approximation, i.e. assume that
\begin{equation}
\frac{R_1-R_2}{R_0} < k(R_1-R_2) \ll 1 ,
\label{eq:thinshell}
\end{equation}
where $k$ is the wavenumber of the perturbations. Note that although shock
1 is relativistic, shock 2 could be either relativistic or nonrelativistic.
In this paper, we only consider the situation where shock 2 is
relativistic.

The equations of motion for an ideal relativistic fluid are
\begin{equation}
\frac{\partial}{\partial t}(\gamma \rho) + {\bf \nabla} \cdot (\gamma \rho
{\bf u}) = 0 ,
\label{eq:mass}
\end{equation}
\begin{equation}
\frac{\gamma^2}{c^2}(e+p)
\left[ \frac{\partial {\bf u}}{\partial t} + ({\bf u} \cdot {\bf \nabla}){\bf u} \right ]
+ \left( {\bf \nabla}p + \frac{{\bf u}}{c^2}\frac{\partial p}{\partial t} \right) = 0 ,
\label{eq:momentum}
\end{equation}
where $\rho$, ${\bf u}$, $e$, $p$ and $\gamma$ are the fluid density,
velocity, energy density, pressure and Lorentz factor, respectively. We
define the surface density $\sigma$, bulk radial velocity $V_r$ and average
tangential velocity ${\bf V}_T$ of region 2 as follows:
\begin{equation}
\sigma(\theta, \phi)=R_0^{-2}\int_{R_0}^{R_1}\gamma \rho r^2 dr ,
\label{eq:sigma}
\end{equation}
\begin{equation}
V_r(\theta, \phi)=(\sigma R_0^2)^{-1}\int_{R_0}^{R_1}\gamma \rho u_r r^2 dr ,
\label{eq:Vr}
\end{equation}
\begin{equation}
{\bf V}_T(\theta, \phi)=(\sigma R_0^2)^{-1}\int_{R_0}^{R_1}\gamma \rho {\bf
u}_T r^2 dr .
\label{eq:VT}
\end{equation}
The time evolution of these variables can be obtained by integrating
equations (\ref{eq:mass}) and (\ref{eq:momentum}) across region 2, using
the boundary conditions at shock 1 and at the contact discontinuity and
neglecting terms of higher order in $(R_1-R_2)/R_0$ and $k(R_1-R_2)$. We
get
\begin{equation}
\partial_t \sigma = -2\frac{V_r}{R_0}\sigma 
+ \rho_1c - \sigma {\bf \nabla}_T \cdot {\bf V}_T ,
\label{eq:dt_sig_5}
\end{equation}
\begin{equation}
\partial_t\gamma(R_0)
= -2\sigma^{-1}\gamma(R_0)\rho_1c
+\frac{3^{3/4}}{2^{1/2}}\sigma^{-1}\frac{p^{3/4}(R_0)\rho_1^{1/4}}
{\gamma^{1/2}(R_0)c^{1/2}}
+\frac{\gamma(R_0)}{2c}V_r{\bf \nabla}_T \cdot {\bf V}_T ,
\label{eq:dtVr7}
\end{equation}
\begin{eqnarray}
\partial_t {\bf V}_T & = & \frac{1}{3}\sigma^{-1}c^2
\left[\rho_1-3^{3/4}\cdot 2^{1/2} \frac{p^{3/4}(R_0)\rho_1^{1/4}}
{\gamma^{3/2}(R_0)c^{3/2}}\right]
{\bf \nabla}_T R_0
-\sigma^{-1}\rho_1c{\bf V}_T
\nonumber\\
& & -\frac{V_r}{R_0}{\bf V}_T 
-\sigma^{-1}\frac{c^2}{3\gamma^2(R_0)}{\bf \nabla}_T \sigma 
+ \frac{c^2}{3\gamma^3(R_0)}{\bf \nabla}_T \gamma(R_0) ,
\label{eq:dtVT5}
\end{eqnarray}
where $\rho_1$ is the density of the unshocked circumburst medium just in
front of shock 1, $\gamma(R_0)$ and $p(R_0)$ are the Lorentz factor and
pressure at the contact discontinuity, and ${\bf \nabla}_T$ denotes the
tangential derivatives. In deriving the above equations we also assumed
that the radial velocities are dominated by the bulk motion of regions 2
and 3 (denoted thereafter as the ``shock layer''), so that $\dot{R}_0
\approx V_r$. The full derivation of the above equations is given in the
Appendix.

From equations (\ref{eq:dt_sig_5}) and (\ref{eq:dtVr7}) we obtain the 
unperturbed equations:
\begin{equation}
\partial_t \sigma^{(0)} = -2\frac{v_c}{R_0^{(0)}}\sigma^{(0)} + \rho_1c ,
\label{eq:unpert_a}
\end{equation}
\begin{equation}
\partial_t \gamma_c = -2(\sigma^{(0)})^{-1} \gamma_c\rho_1 c
+\frac{3^{3/4}}{2^{1/2}}(\sigma^{(0)})^{-1}
\frac{p_c^{3/4}\rho_1^{1/4}}{\gamma_c^{1/2}c^{1/2}} ,
\label{eq:unpert_b}
\end{equation}
where $v_c=\dot{R}_0^{(0)} = V_r^{(0)}$ and $\gamma_c = \gamma^{(0)}(R_0) =
1/\sqrt{1-v_c^2/c^2}$ are the velocity and Lorentz factor of the
unperturbed contact discontinuity, $p_c=p^{(0)}(R_0)$ is the unperturbed
pressure at the contact discontinuity, and we use a superscript (0) to
denote unperturbed values.

For the shocked shell in region 3, we define the surface density, bulk
radial velocity and average tangential velocity to be:
\begin{equation}
\sigma_3(\theta, \phi)=R_0^{-2}\int_{R_2}^{R_0}\gamma \rho r^2 dr ,
\label{eq:sigma_shell}
\end{equation}
\begin{equation}
V_{r3}(\theta, \phi)=(\sigma_3 R_0^2)^{-1} 
\int_{R_2}^{R_0}\gamma \rho u_r r^2 dr ,
\label{eq:Vr_shell}
\end{equation}
\begin{equation}
{\bf V}_{T3}(\theta, \phi)=(\sigma_3 R_0^2)^{-1}
\int_{R_2}^{R_0}\gamma \rho {\bf u}_T r^2 dr ,
\label{eq:VT_shell}
\end{equation}
where a subscript 3 denotes quantities in region 3. The time derivatives of
the above variables can be derived similarly to those in region 2,
\begin{equation}
\partial_t \sigma_3 = -2\frac{V_r}{R_0}\sigma_3 
+ \frac{\gamma_4}{\gamma^2(R_0)}\rho_4c 
- \sigma_3 {\bf \nabla}_T \cdot {\bf V}_{T3} ,
\label{eq:dt_sig_shell}
\end{equation}
\begin{equation}
\partial_t\gamma(R_0) = \sigma_3^{-1}\frac{\gamma_4}{\gamma(R_0)}\rho_4c
-3^{3/4}\sigma_3^{-1}\frac{p^{3/4}(R_0)\rho_4^{1/4}\gamma^{1/2}(R_0)}
{\gamma_4^{1/2}c^{1/2}} +\frac{\gamma(R_0)}{2c}V_{r}{\bf \nabla}_T \cdot
{\bf V}_{T3} ,
\label{eq:dtVr_shell}
\end{equation}
\begin{eqnarray} 
\partial_t {\bf V}_{T3} &=& \frac{1}{3}\sigma_3^{-1}c^2 \left[2\cdot
3^{3/4} \frac{p^{3/4}(R_0)\rho_4^{1/4}}
{\gamma_4^{1/2}\gamma^{1/2}(R_0)c^{3/2}}
-\frac{\gamma_4}{2\gamma^2(R_0)}\rho_4\right] {\bf \nabla}_T R_0
\nonumber\\ && -\sigma_3^{-1}\frac{\gamma_4}{\gamma^2(R_0)}\rho_4c{\bf
V}_{T3} -\frac{V_r}{R_0}{\bf V}_{T3} -\frac{1}{4\gamma^2(R_0)}
\left(\frac{\partial_t\rho_4}{\rho_4}\right){\bf V}_{T3} \nonumber\\ &&
-\sigma_3^{-1}\frac{c^2}{3\gamma^2(R_0)}{\bf \nabla}_T \sigma_3 +
\frac{c^2}{3\gamma^3(R_0)}{\bf \nabla}_T \gamma(R_0) ,
\label{eq:dtVT5_shell}
\end{eqnarray}
where $\rho_4$ and $\gamma_4$ are the density and Lorentz factor of the
unshocked shell (region 4) just in front of shock 2. We have also used the
relation $V_{r3} \approx V_r$, as appropriate in the thin shell
approximation.

The unperturbed equations for region 3 are 
\begin{equation}
\partial_t \sigma_3^{(0)} = -2\frac{v_c}{R_0^{(0)}}\sigma_3^{(0)} 
+ \frac{\gamma_4}{\gamma_c^2}\rho_4c 
\label{eq:unpert_a_3}
\end{equation}
\begin{equation}
\partial_t \gamma_c = (\sigma_3^{(0)})^{-1}
\frac{\gamma_4}{\gamma_c}\rho_4c -3^{3/4}(\sigma_3^{(0)})^{-1}
\frac{p_c^{3/4}\rho_4^{1/4} \gamma_c^{1/2}} {\gamma_4^{1/2}c^{1/2}}
\label{eq:unpert_b_3}
\end{equation}

Equations (\ref{eq:dt_sig_5}), (\ref{eq:dtVr7}), (\ref{eq:dtVT5}),
(\ref{eq:dt_sig_shell}), (\ref{eq:dtVr_shell}) and (\ref{eq:dtVT5_shell})
make a complete set of equations for the evolution of the forward/reverse 
shock system. The perturbation equations depend on the density
profile of the circumburst medium.  In the following subsections, we shall
derive the perturbation equations for a uniform medium (such as the ISM)
and for a progenitor wind.

\subsection{Uniform Medium (ISM)}

For a uniform circumburst medium, $\rho_1={\rm const}$.  We define the
perturbation variables $\delta \equiv \sigma / \sigma^{(0)} - 1$, $\delta_3
\equiv \sigma_3 / \sigma_3^{(0)} - 1$, $\Delta R \equiv R_0 - R_0^{(0)}$
and $\Delta_p \equiv p(R_0) / p^{(0)}(R_0) - 1$.  From equations
(\ref{eq:dt_sig_5}), (\ref{eq:dtVr7}) and (\ref{eq:dtVT5}), we obtain the
following perturbation equations:
\begin{equation}
\partial_t \delta = - \frac{2}{R_0^{(0)}}\partial_t \Delta R 
+ 2 \frac{v_c}{(R_0^{(0)})^2}\Delta R - (\sigma^{(0)})^{-1}\rho_1 c \delta 
- {\bf \nabla}_T \cdot {\bf V}_T ,
\label{eq:pert_a}
\end{equation}
\begin{eqnarray}
\partial_t^2 \Delta R & = &
\left[2(\sigma^{(0)})^{-1}\frac{\rho_1c^2}{\gamma_c^2}
-\frac{3^{3/4}}{2^{1/2}}(\sigma^{(0)})^{-1}
\frac{p_c^{3/4}\rho_1^{1/4}c^{1/2}}{\gamma_c^{7/2}}\right]\delta
\nonumber\\ & & + \left[4(\sigma^{(0)})^{-1}\rho_1c -\frac{7\cdot
3^{3/4}}{2^{3/2}}(\sigma^{(0)})^{-1}
\frac{p_c^{3/4}\rho_1^{1/4}}{\gamma_c^{3/2}c^{1/2}} \right]\partial_t
\Delta R \nonumber\\ & & +\frac{3^{7/4}}{2^{5/2}}(\sigma^{(0)})^{-1}
\frac{p_c^{3/4}\rho_1^{1/4}c^{1/2}}{\gamma_c^{7/2}}\Delta_p
+\frac{v_c}{2\gamma_c^2}{\bf \nabla}_T \cdot {\bf V}_T ,
\label{eq:pert_b}
\end{eqnarray}
\begin{eqnarray}
\partial_t {\bf V}_T &=& \frac{1}{3}(\sigma^{(0)})^{-1}c^2
\left[\rho_1-3^{3/4}\cdot 2^{1/2} \frac{p_c^{3/4}\rho_1^{1/4}}
{\gamma_c^{3/2}c^{3/2}}\right] {\bf \nabla}_T \Delta R
-(\sigma^{(0)})^{-1}\rho_1c{\bf V}_T \nonumber\\ & &
-\frac{v_c}{R_0^{(0)}}{\bf V}_T -\frac{c^2}{3\gamma_c^2}{\bf \nabla}_T
\delta + \frac{c}{3} {\bf \nabla}_T (\partial_t \Delta R) .
\label{eq:pert_c}
\end{eqnarray}

Assuming that $\gamma_4$ is a constant and there is no shell spreading, we
get $\rho_4 \propto R^{-2}$. Using this scaling we derive the following
perturbation equations from equations (\ref{eq:dt_sig_shell}),
(\ref{eq:dtVr_shell}) and (\ref{eq:dtVT5_shell}):
\begin{eqnarray}
\partial_t \delta_3 & = & -2(\sigma_3^{(0)})^{-1}\gamma_4\rho_4\partial_t
\Delta R + \left[ 2 \frac{v_c}{(R_0^{(0)})^2} - 2
(\sigma_3^{(0)})^{-1}\frac{\gamma_4}{\gamma_c^2} \frac{\rho_4 c}{R_0^{(0)}}
\right] \Delta R \nonumber\\ & & -
(\sigma_3^{(0)})^{-1}\frac{\gamma_4}{\gamma_c^2}\rho_4 c \delta_3 - {\bf
\nabla}_T \cdot {\bf V}_{T3} ,
\label{eq:pert_a_3}
\end{eqnarray}
\begin{eqnarray}
\partial_t^2 \Delta R & = &
\left[-(\sigma_3^{(0)})^{-1}\frac{\gamma_4}{\gamma_c^4}\rho_4c^2
+3^{3/4}(\sigma_3^{(0)})^{-1}
\frac{p_c^{3/4}\rho_4^{1/4}c^{1/2}}{\gamma_4^{1/2}\gamma_c^{5/2}}\right]\delta_3
\nonumber\\ & & +
\left[-4(\sigma_3^{(0)})^{-1}\frac{\gamma_4}{\gamma_c^2}\rho_4c
+\frac{5\cdot 3^{3/4}}{2}(\sigma_3^{(0)})^{-1}
\frac{p_c^{3/4}\rho_4^{1/4}}{\gamma_4^{1/2}\gamma_c^{1/2}c^{1/2}}
\right]\partial_t \Delta R \nonumber\\ & & +
\left[-2(\sigma_3^{(0)})^{-1}\frac{\gamma_4}{\gamma_c^4}\frac{\rho_4c^2}{R_0^{(0)}}
+\frac{3^{3/4}}{2}(\sigma_3^{(0)})^{-1}
\frac{p_c^{3/4}\rho_4^{1/4}c^{1/2}}{\gamma_4^{1/2}\gamma_c^{5/2}}
\frac{1}{R_0^{(0)}}\right]\Delta R \nonumber\\ & &
-\frac{3^{7/4}}{4}(\sigma_3^{(0)})^{-1}
\frac{p_c^{3/4}\rho_4^{1/4}c^{1/2}}{\gamma_4^{1/2}\gamma_c^{5/2}}\Delta_p
+\frac{v_c}{2\gamma_c^2}{\bf \nabla}_T \cdot {\bf V}_{T3} ,
\label{eq:pert_b_3}
\end{eqnarray}
\begin{eqnarray} 
\partial_t {\bf V}_{T3} &=& \frac{1}{3}(\sigma_3^{(0)})^{-1}c^2
\left[2\cdot 3^{3/4} \frac{p_c^{3/4}\rho_4^{1/4}}
{\gamma_4^{1/2}\gamma_c^{1/2}c^{3/2}}
-\frac{\gamma_4}{2\gamma_c^2}\rho_4\right] {\bf \nabla}_T \Delta R
-(\sigma_3^{(0)})^{-1}\frac{\gamma_4}{\gamma_c^2}\rho_4c{\bf V}_{T3}
\nonumber\\ && -\frac{v_c}{R_0^{(0)}}{\bf V}_{T3} -\frac{1}{4\gamma_c^2}
\left(\frac{\partial_t\rho_4}{\rho_4}\right){\bf V}_{T3}
-\frac{c^2}{3\gamma_c^2}{\bf \nabla}_T \delta_3 + \frac{c}{3} {\bf
\nabla}_T (\partial_t \Delta R) .
\label{eq:pert_c_3}
\end{eqnarray}

In total, we have six perturbation equations
(\ref{eq:pert_a})--(\ref{eq:pert_c_3}) in six variables: $\delta$,
$\delta_3$, $\Delta R$, ${\bf V}_T$, ${\bf V}_{T3}$ and $\Delta_p$.  In
order to solve these equations, we first need to find the unperturbed
values $\sigma^{(0)}$, $\sigma_3^{(0)}$, $\gamma_c$ and $p_c$ from
equations (\ref{eq:unpert_a}), (\ref{eq:unpert_b}), (\ref{eq:unpert_a_3})
and (\ref{eq:unpert_b_3}).  The time dependence of $\gamma_c$ has been
derived by Sari \& Piran (1995) and Sari et al. (1996). When both the
forward shock and the reverse shock are ultrarelativistic and strong,
\begin{equation}
\gamma_c \propto \gamma_4^{1/2}f^{1/4} ,
\label{eq:gammac}
\end{equation}
where $f=\rho_4/\rho_1$. For $\rho_1={\rm const}$, 
$\gamma_4={\rm const}$ and $\rho_4 \propto R^{-2}$, we get 
\begin{equation}
\gamma_c \propto R^{-1/2} \propto t^{-1/2} .
\label{eq:gammac_t}
\end{equation}

With $v_c \approx c$ and $R_0^{(0)} \approx ct$, equation
(\ref{eq:unpert_a}) yields
\begin{equation}
\sigma^{(0)} \approx \frac{1}{3}\rho_1 ct.
\label{eq:sigma_sol}
\end{equation}
For $\rho_4 \propto t^{-2}$ and $\gamma_c \propto t^{-1/2}$, 
equation (\ref{eq:unpert_a_3}) gives
\begin{equation}
\sigma_3^{(0)} \approx \frac{1}{2} \frac{\gamma_4}{\gamma_c^2} \rho_4 ct
=const .
\label{eq:sigma3_sol}
\end{equation}
By substituting equation (\ref{eq:sigma_sol}) into equation 
(\ref{eq:unpert_b}) we find
\begin{equation}
p_c \approx \frac{11^{4/3}}{3^{7/3}\cdot 2^{2/3}}\gamma_c^2 \rho_1 c^2
= 1.187 \gamma_c^2 \rho_1 c^2 < \frac{4}{3}\gamma_c^2 \rho_1 c^2 ,
\label{eq:pc_sol}
\end{equation}
where $(4/3)\gamma_c^2 \rho_1 c^2$ is the pressure just behind the forward
shock. Similarly, by substituting equation (\ref{eq:sigma3_sol}) into
equation (\ref{eq:unpert_b_3}) we get
\begin{equation}
p_c \approx \frac{5^{4/3}}{4^{4/3}\cdot
3}\frac{\gamma_4^2}{\gamma_c^2}\rho_4c^2
=0.449\frac{\gamma_4^2}{\gamma_c^2}\rho_4c^2 > \frac{4}{3}
\bar{\gamma_3}^2\rho_4c^2 ,
\label{eq:pc_sol_3}
\end{equation}
where $\bar{\gamma_3} \approx \gamma_4/(2\gamma_c)$ is the Lorentz factor
of the shocked shell (region 3) with respect to the unshocked shell (region
4) and $(4/3)\bar{\gamma_3}^2\rho_4c^2$ is the pressure just behind the
reverse shock. The pressure difference between the two sides of the layer
causes it to decelerate. By combining equations (\ref{eq:pc_sol}) and
(\ref{eq:pc_sol_3}) we find
\begin{equation}
\gamma_c \approx 0.784 \gamma_4^{1/2}(\rho_4/\rho_1)^{1/4} .
\label{eq:gamma_c_new}
\end{equation}

Substitution of the values of $\sigma^{(0)}$, $\sigma_3^{(0)}$ and $p_c$
into the perturbation equations (\ref{eq:pert_a})--(\ref{eq:pert_c_3})
yields, 
\begin{equation}
\partial_t \delta = -\frac{2}{ct}\partial_t\Delta R + \frac{2}{ct^2}\Delta R
-\frac{3}{t}\delta - {\bf \nabla}_T \cdot {\bf V}_T ,
\label{eq:pert_delta}
\end{equation}
\begin{equation}
\partial_t^2 \Delta R = \frac{1}{2}\frac{c}{\gamma_c^2t}\delta
- \frac{29}{4}\frac{1}{t}\partial_t \Delta R 
+ \frac{33}{8}\frac{c}{\gamma_c^2t}\Delta_p
+ \frac{c}{2\gamma_c^2}{\bf \nabla}_T \cdot {\bf V}_T ,
\label{eq:pert_DeltaR}
\end{equation}
\begin{equation}
\partial_t {\bf V}_T = -\frac{8}{3}\frac{c}{t}{\bf \nabla}_T \Delta R 
-\frac{4}{t}{\bf V}_T -\frac{c^2}{3\gamma_c^2}{\bf \nabla}_T \delta
+ \frac{c}{3} {\bf \nabla}_T (\partial_t \Delta R) ,
\label{eq:pert_VT}
\end{equation}
\begin{equation}
\partial_t \delta_3 = -4\frac{\gamma_c^2}{ct}\partial_t\Delta R - \frac{2}{ct^2}\Delta R
-\frac{2}{t}\delta_3 - {\bf \nabla}_T \cdot {\bf V}_{T3} ,
\label{eq:pert_delta3}
\end{equation}
\begin{equation}
\partial_t^2 \Delta R = \frac{1}{2}\frac{c}{\gamma_c^2t}\delta_3 -
\frac{7}{4}\frac{1}{t}\partial_t \Delta R -\frac{11}{4}\frac{1}{\gamma_c^2
t^2}\Delta R - \frac{15}{8}\frac{c}{\gamma_c^2t}\Delta_p +
\frac{c}{2\gamma_c^2}{\bf \nabla}_T \cdot {\bf V}_{T3} ,
\label{eq:pert_DeltaR3}
\end{equation}
\begin{equation}
\partial_t {\bf V}_{T3} = \frac{4}{3}\frac{c}{t}{\bf \nabla}_T \Delta R 
-\frac{3}{t}{\bf V}_{T3} -\frac{c^2}{3\gamma_c^2}{\bf \nabla}_T \delta_3
+ \frac{c}{3} {\bf \nabla}_T (\partial_t \Delta R) .
\label{eq:pert_VT3}
\end{equation}
Combining equations (\ref{eq:pert_DeltaR}) and (\ref{eq:pert_DeltaR3}) and
eliminating $\Delta_p$, we get
\begin{eqnarray}
\partial_t^2 \Delta R & = &  \frac{5}{32}\frac{c}{\gamma_c^2t}\delta
+ \frac{11}{32}\frac{c}{\gamma_c^2t}\delta_3
- \frac{111}{32}\frac{1}{t}\partial_t \Delta R
- \frac{121}{64}\frac{1}{\gamma_c^2 t^2}\Delta R
\nonumber\\
& & 
+ \frac{5}{32}\frac{c}{\gamma_c^2}{\bf \nabla}_T \cdot {\bf V}_{T}
+ \frac{11}{32}\frac{c}{\gamma_c^2}{\bf \nabla}_T \cdot {\bf V}_{T3} .
\label{eq:pert_DeltaRnew}
\end{eqnarray}

Next we expand the spatial dependence of the perturbation variables in
spherical harmonics. We choose to normalize these variables so as to make
them dimensionless and of a similar magnitude through the definitions
\begin{equation}
\Delta R = \sum_{l,m} \Delta R(l,m,t)
[R_0^{(0)}/\gamma_c^2]Y_{lm}(\theta,\phi) ,
\label{eq:DeltaR_spher}
\end{equation}  
\begin{equation}
\delta = \sum_{l,m}\delta (l,m,t)Y_{lm}(\theta,\phi) ,
\label{eq:delta_spher}
\end{equation}
\begin{equation}
\delta_3 = \sum_{l,m} \delta_3(l,m,t)Y_{lm}(\theta,\phi) ,
\label{eq:delta3_spher}
\end{equation}
\begin{equation}
{\bf V}_T = \sum_{l,m} V_T(l,m,t)[cR_0^{(0)}/\gamma_c]{{\bf
\nabla}_TY_{lm}(\theta,\phi) \over l} ,
\label{eq:VT_spher}
\end{equation}
\begin{equation}
{\bf V}_{T3} = \sum_{l,m} V_{T3}(l,m,t)[cR_0^{(0)}/\gamma_c]{{\bf
\nabla}_TY_{lm}(\theta,\phi)\over l} .
\label{eq:VT3_spher}
\end{equation}

Equations (\ref{eq:pert_delta}), (\ref{eq:pert_VT}),
(\ref{eq:pert_delta3}), (\ref{eq:pert_VT3}) and (\ref{eq:pert_DeltaRnew})
can be rewritten as
\begin{equation}
d_t \Delta R = F ,
\label{eq:pert_DeltaR_a}
\end{equation}
\begin{equation}
d_t \delta = -\frac{2}{\gamma_c^2}F -\frac{2}{\gamma_c^2t}\Delta R
-\frac{3}{t}\delta +\frac{(l+1)}{\gamma_c t}V_T ,
\label{eq:pert_delta_a}
\end{equation}
\begin{equation}
d_t V_T = \frac{l}{3\gamma_c}F
-\frac{2l}{\gamma_c t}\Delta R
-\frac{l}{3\gamma_c t}\delta -\frac{9}{2t}V_T,
\label{eq:pert_VT_a}
\end{equation}
\begin{equation}
d_t \delta_3 = -4F -\frac{8}{t}\Delta R
-\frac{2}{t}\delta_3 +\frac{(l+1)}{\gamma_c t}V_{T3} ,
\label{eq:pert_delta3_a}
\end{equation}
\begin{equation}
d_t V_{T3} = \frac{l}{3\gamma_c}F
+\frac{2l}{\gamma_c t}\Delta R
-\frac{l}{3\gamma_c t}\delta_3 -\frac{7}{2t}V_{T3},
\label{eq:pert_VT3_a}
\end{equation}
\begin{equation}
d_t F = 
- \frac{239}{32t}F - \frac{143}{16t^2}\Delta R
+ \frac{5}{32t^2}\delta + \frac{11}{32t^2}\delta_3
- \frac{5}{32}\frac{(l+1)}{\gamma_c t^2}V_T
- \frac{11}{32}\frac{(l+1)}{\gamma_c t^2}V_{T3} .
\label{eq:pert_F_a}
\end{equation}
The perturbation variables in the above six equations are dimensionless and
only functions of time.  We have added the variable $F$ so that all the
equations will have the form of first order differential equations.

\subsection{Wind Medium}

If the circumburst medium is a progenitor wind, $\rho_1 \propto
R^{-2}$. Accordingly, the perturbation equations (\ref{eq:pert_a}) and
(\ref{eq:pert_b}) need to be changed to
\begin{equation}
\partial_t \delta = - \frac{2}{R_0^{(0)}}\partial_t \Delta R 
+ \left[ 2\frac{v_c}{(R_0^{(0)})^2} 
- 2(\sigma^{(0)})^{-1} \frac{\rho_1c}{R_0^{(0)}}\right]\Delta R 
- (\sigma^{(0)})^{-1}\rho_1 c \delta 
- {\bf \nabla}_T \cdot {\bf V}_T ,
\label{eq:pert_a_wind}
\end{equation}
\begin{eqnarray}
\partial_t^2 \Delta R & = &
\left[2(\sigma^{(0)})^{-1}\frac{\rho_1c^2}{\gamma_c^2}
-\frac{3^{3/4}}{2^{1/2}}(\sigma^{(0)})^{-1}
\frac{p_c^{3/4}\rho_1^{1/4}c^{1/2}}{\gamma_c^{7/2}}\right]\delta
\nonumber\\ & & + \left[4(\sigma^{(0)})^{-1}\rho_1c -\frac{7\cdot
3^{3/4}}{2^{3/2}}(\sigma^{(0)})^{-1}
\frac{p_c^{3/4}\rho_1^{1/4}}{\gamma_c^{3/2}c^{1/2}} \right]\partial_t
\Delta R \nonumber\\ & & +
\left[4(\sigma^{(0)})^{-1}\frac{\rho_1c^2}{\gamma_c^2 R_0^{(0)}}
-\frac{3^{3/4}}{2^{3/2}}(\sigma^{(0)})^{-1}
\frac{p_c^{3/4}\rho_1^{1/4}c^{1/2}}{\gamma_c^{7/2} R_0^{(0)}}\right]\Delta
R \nonumber\\ & & +\frac{3^{7/4}}{2^{5/2}}(\sigma^{(0)})^{-1}
\frac{p_c^{3/4}\rho_1^{1/4}c^{1/2}}{\gamma_c^{7/2}}\Delta_p
+\frac{v_c}{2\gamma_c^2}{\bf \nabla}_T \cdot {\bf V}_T .
\label{eq:pert_b_wind}
\end{eqnarray}
Equations (\ref{eq:pert_c})--(\ref{eq:pert_c_3}) remain the same
as in the uniform medium case.

Since $\rho_1 \propto R^{-2}$ and $\rho_4 \propto R^{-2}$ in the wind case,
equation (\ref{eq:gammac}) implies that $\gamma_c$ is constant over
time. Equations (\ref{eq:unpert_a}), (\ref{eq:unpert_b}),
(\ref{eq:unpert_a_3}) and (\ref{eq:unpert_b_3}) then yield the unperturbed
parameters:
\begin{equation}
\sigma^{(0)} \approx \rho_1ct,
\label{eq:sigma_wind}
\end{equation}
\begin{equation}
\sigma^{(0)}_3 \approx \frac{\gamma_4}{\gamma_c^2}\rho_4ct,
\label{eq:sigma3_wind}
\end{equation}
\begin{equation}
p_c \approx \frac{4}{3}\gamma_c^2\rho_1c^2 
\approx \frac{1}{3}\frac{\gamma_4^2}{\gamma_c^2}\rho_4c^2,
\label{eq:pc_wind}
\end{equation}
\begin{equation}
\gamma_c \approx \frac{1}{\sqrt{2}}\gamma_4^{1/2}
\left(\frac{\rho_4}{\rho_1}\right)^{1/4} .
\label{eq:gammac_wind}
\end{equation}

Substitution of equations (\ref{eq:sigma_wind})--(\ref{eq:pc_wind}) into
the perturbation equations (\ref{eq:pert_a_wind}), (\ref{eq:pert_b_wind})
and (\ref{eq:pert_c})--(\ref{eq:pert_c_3}), yields
\begin{equation}
\partial_t \delta = -\frac{2}{ct}\partial_t\Delta R 
- \frac{1}{t}\delta - {\bf \nabla}_T \cdot {\bf V}_T ,
\label{eq:pert_delta_wind}
\end{equation}
\begin{equation}
\partial_t^2 \Delta R = 
- \frac{3}{t}\partial_t \Delta R 
+ \frac{3}{\gamma_c^2t^2}\Delta R 
+ \frac{3}{2}\frac{c}{\gamma_c^2t}\Delta_p
+ \frac{c}{2\gamma_c^2}{\bf \nabla}_T \cdot {\bf V}_T ,
\label{eq:pert_DeltaR_wind}
\end{equation}
\begin{equation}
\partial_t {\bf V}_T = -\frac{c}{t}{\bf \nabla}_T \Delta R 
-\frac{2}{t}{\bf V}_T -\frac{c^2}{3\gamma_c^2}{\bf \nabla}_T \delta
+ \frac{c}{3} {\bf \nabla}_T (\partial_t \Delta R) ,
\label{eq:pert_VT_wind}
\end{equation}
\begin{equation}
\partial_t \delta_3 = -2\frac{\gamma_c^2}{ct}\partial_t\Delta R 
-\frac{1}{t}\delta_3 - {\bf \nabla}_T \cdot {\bf V}_{T3} ,
\label{eq:pert_delta3_wind}
\end{equation}
\begin{equation}
\partial_t^2 \Delta R = 
- \frac{3}{2}\frac{1}{t}\partial_t \Delta R
-\frac{3}{2}\frac{1}{\gamma_c^2 t^2}\Delta R 
- \frac{3}{4}\frac{c}{\gamma_c^2t}\Delta_p
+ \frac{c}{2\gamma_c^2}{\bf \nabla}_T \cdot {\bf V}_{T3} ,
\label{eq:pert_DeltaR3_wind}
\end{equation}
\begin{equation}
\partial_t {\bf V}_{T3} = \frac{1}{2}\frac{c}{t}{\bf \nabla}_T \Delta R 
-\frac{2}{t}{\bf V}_{T3} -\frac{c^2}{3\gamma_c^2}{\bf \nabla}_T \delta_3
+ \frac{c}{3} {\bf \nabla}_T (\partial_t \Delta R) .
\label{eq:pert_VT3_wind}
\end{equation}
Combining equations (\ref{eq:pert_DeltaR_wind}) and
(\ref{eq:pert_DeltaR3_wind}), and eliminating $\Delta_p$, we get
\begin{equation}
\partial_t^2 \Delta R = 
- \frac{2}{t}\partial_t \Delta R
+ \frac{c}{6\gamma_c^2}{\bf \nabla}_T \cdot {\bf V}_{T}
+ \frac{c}{3\gamma_c^2}{\bf \nabla}_T \cdot {\bf V}_{T3} .
\label{eq:pert_DeltaRnew_wind}
\end{equation}

Using the same normalized perturbation variables as defined in equations
(\ref{eq:DeltaR_spher})--(\ref{eq:VT3_spher}), we can rewrite equations
(\ref{eq:pert_delta_wind}), (\ref{eq:pert_VT_wind}),
(\ref{eq:pert_delta3_wind}), (\ref{eq:pert_VT3_wind}) and
(\ref{eq:pert_DeltaRnew_wind}) as
\begin{equation}
d_t \Delta R = F
\label{eq:pert_DeltaR_a_wind} ,
\end{equation}
\begin{equation}
d_t \delta = -\frac{2}{\gamma_c^2}F -\frac{2}{\gamma_c^2t}\Delta R
-\frac{1}{t}\delta +\frac{(l+1)}{\gamma_c t}V_T ,
\label{eq:pert_delta_a_wind}
\end{equation}
\begin{equation}
d_t V_T = \frac{l}{3\gamma_c}F
-\frac{2l}{3\gamma_c t}\Delta R 
-\frac{l}{3\gamma_c t}\delta -\frac{2}{t}V_T ,
\label{eq:pert_VT_a_wind}
\end{equation}
\begin{equation}
d_t \delta_3 = -2F -\frac{2}{t}\Delta R
-\frac{1}{t}\delta_3 +\frac{(l+1)}{\gamma_c t}V_{T3} ,
\label{eq:pert_delta3_a_wind}
\end{equation}
\begin{equation}
d_t V_{T3} = \frac{l}{3\gamma_c}F
+ \frac{5l}{6\gamma_c t}\Delta R 
-\frac{l}{3\gamma_c t}\delta_3 -\frac{2}{t}V_{T3} ,
\label{eq:pert_VT3_a_wind}
\end{equation}
\begin{equation}
d_t F = - \frac{4}{t}F - \frac{2}{t^2}\Delta R
- \frac{(l+1)}{6 \gamma_c t^2}V_T
- \frac{(l+1)}{3 \gamma_c t^2}V_{T3} .
\label{eq:pert_F_a_wind}
\end{equation}
These six first order differential equations are the final perturbation
equations for the wind case.

\section{Solutions of the Perturbations Equations}

For the wind case, $\gamma_c={\rm const}$, and we can solve the
perturbation equations analytically. If we define $F=F'/t$, then equations
(\ref{eq:pert_DeltaR_a_wind})--(\ref{eq:pert_F_a_wind}) can be rewritten in
a matrix form:
\begin{equation}
d_t \left(\begin{array}{c}
F' \\ \Delta R \\ \delta \\ V_T \\ \delta_3 \\ V_{T3}
\end{array} \right)
= \frac{1}{t} \left(\begin{array}{cccccc} -3 & -2 & 0 &
-\frac{(l+1)}{6\gamma_c} & 0 & -\frac{(l+1)}{3\gamma_c} \\ 1 & 0 & 0 & 0 &
0 & 0 \\ -\frac{2}{\gamma_c^2} & -\frac{2}{\gamma_c^2} & -1 &
\frac{(l+1)}{\gamma_c} & 0 & 0 \\ \frac{l}{3\gamma_c} &
-\frac{2l}{3\gamma_c} & -\frac{l}{3\gamma_c} & -2 & 0 & 0 \\ -2 & -2 & 0 &
0 & -1 & \frac{(l+1)}{\gamma_c} \\ \frac{l}{3\gamma_c} &
\frac{5l}{6\gamma_c} & 0 & 0 & -\frac{l}{3\gamma_c} & -2
\end{array} \right)
\left(\begin{array}{c}
F' \\ \Delta R \\ \delta \\ V_T \\ \delta_3 \\ V_{T3}
\end{array} \right) .
\label{eq:windmatrix}
\end{equation}
In matrix notation, the above equation is
\begin{equation}
d_t{\bf y} = \frac{1}{t} {\bf A} \cdot {\bf y} ,
\label{eq:matrixeqn1}
\end{equation}
where ${\bf y}$ is a vector, and ${\bf A}$ is a $6 \times 6$
time-independent matrix.

The matrix ${\bf A}$ can be diagonalized through the transformation
\begin{equation}
{\bf X}^{-1} \cdot {\bf A} \cdot {\bf X} = 
{\rm diag} (\lambda_1 \cdots \lambda_6) = {\bf D} ,
\label{eq:matrixdiag}
\end{equation}
where $\lambda_1$, $\cdots$, $\lambda_6$ are the eigenvalues of the matrix
{\bf A}, and ${\bf X}$ is the matrix formed by columns from the
eigenvectors (i.e. the $k$-th column of ${\bf X}$ is the eigenvector
corresponding to the eigenvalue $\lambda_k$). Equation
(\ref{eq:matrixeqn1}) can then be transformed to
\begin{equation}
d_t{\bf y} = \frac{1}{t} ({\bf X} \cdot {\bf D} \cdot {\bf X}^{-1}) \cdot
{\bf y} \Longrightarrow d_t({\bf X}^{-1} \cdot {\bf y}) = \frac{1}{t} {\bf
D} \cdot ({\bf X}^{-1} \cdot {\bf y}) .
\label{eq:matrixeqn2}
\end{equation}
By defining a new vector ${\bf y}'={\bf X}^{-1} \cdot {\bf y}$, we get
the equation
\begin{equation}
d_t{\bf y}' = \frac{1}{t} {\bf D} \cdot {\bf y}' ,
\label{eq:matrixeqn3}
\end{equation}
which has six components
\begin{equation}
d_t y'_k = \frac{1}{t} \lambda_k y'_k, \ \ \ \ k =1, \cdots, 6.
\label{eq:matrixeqnk}
\end{equation}
Since $\lambda_k$ can be a complex number, we write $\lambda_k = a_k +
ib_k$. The solution to equation (\ref{eq:matrixeqnk}) is then
\begin{equation}
y'_k = c_k t^{a_k + ib_k} = c_k t^{a_k}e^{ib_k \ln t} ,
\label{eq:ykp}
\end{equation}
where $c_k$ is a constant dictated by the initial conditions.  Each $y'_k$
defines a mode of the shock system.  There are six modes in total
corresponding to six eigenvalues of the matrix ${\bf A}$. The real part of
each eigenvalue dictates the overall temporal behavior of the mode, while
the imaginary part determines its oscillation frequency. The vector ${\bf
y}$ can be derived from the relation
\begin{equation}
{\bf y} = {\bf X} \cdot {\bf y}' .
\label{eq:ysol}
\end{equation}
Hence, each component of the vector ${\bf y}$ is a linear combination of
the six different modes.  The vector ${\bf c}$ whose components are $c_k$,
can be obtained from the initial conditions
\begin{equation}
{\bf c} = {\bf y}'(t=1) = {\bf X}^{-1} \cdot {\bf y}(t=1) ,
\label{eq:cinit}
\end{equation}
namely
\begin{equation}
\left(\begin{array}{c}
c_1 \\ c_2 \\ c_3 \\ c_4 \\ c_5 \\ c_6
\end{array} \right)
= {\bf X}^{-1} \cdot \left(\begin{array}{c} F'(t=1) \\ \Delta R(t=1) \\
\delta(t=1) \\ V_T(t=1) \\ \delta_3(t=1) \\ V_{T3}(t=1)
\end{array} \right) .
\label{eq:cinitlist}
\end{equation}

The eigenvalues of the matrix ${\bf A}$ can be calculated for different
values of $\gamma_c$ and $l$. Figure 2 shows the real and imaginary parts
of the six eigenvalues as functions of $l$ for $\gamma_c = 500$. Each row
in the figure contains two panels, corresponding to the real and imaginary
parts of a particular eigenvalue. The results show that for $l \le 320$,
all the six eigenvalues are real numbers, and so there are no oscillations.
For $ 320 < l \le 432$, two eigenvalues are complex numbers and they are
a pair of complex conjugates, implying that two modes are oscillating with
the same frequency. For $l > 432$, there are two pairs of complex
conjugates. The transition from real eigenvalues to complex eigenvalues
occurs when $l \sim \gamma_c$, as expected from the fact that oscillations
are possible only when causality allows communication across the scale of a
wavelength for modes with $l \ga \gamma_c$.

For the thin shell approximation to be valid, we require that the
wavelength of the perturbation be much larger than the thickness of the
forward/reverse shock system in the shock frame.  The thickness of the
shock system is $\la 2 R_0/\zeta \gamma_c^2$ in the observer frame, and
thus $\la 2 R_0/\zeta \gamma_c$ in the shock frame.  Here $\zeta$ is a
constant that ranges between $\sim 4$ and $\sim 12$ for the wind and ISM
profiles, respectively. 
The wavelength of the perturbation is $\sim 2\pi R_0/l$ in the shock frame.
Therefore, we enforce an upper limit on $l$ of $\sim 10\gamma_c$.  Figure 2
shows that for all values of $l$, the real parts of the six eigenvalues are
$\leq -1$. This implies that all modes are decaying faster or proportional
to $t^{-1}$.  Since each perturbation variable is a linear combination of
the six modes, we conclude that all perturbation variables should also
decay faster than or equal to $t^{-1}$. Thus the system is stable.  Note
that for large $\gamma_c$ we expect the results to depend only on
$l/\gamma_c$ [see equation (\ref{eq:windmatrix})], and so our particular
choice of $\gamma_c=500$ can be scaled appropriately to other values of
$\gamma_c$.

For $l \gg \gamma_c$ the eigenvalues admit the following analytical
solutions,
\begin{equation}
\lambda_1 = -1 ,
\ \ \ 
\lambda_2 = -\frac{3}{2} - \frac{1}{\sqrt{3}}\frac{l}{\gamma_c}i ,
\ \ \ 
\lambda_3 = -\frac{3}{2} + \frac{1}{\sqrt{3}}\frac{l}{\gamma_c}i ,
\label{eq:eigenvallargel1}
\end{equation}
\begin{equation}
\lambda_4 = -\frac{19}{9} ,
\ \ \ 
\lambda_5 = -\frac{13}{9} + \frac{1}{\sqrt{2}}\frac{l}{\gamma_c}i ,
\ \ \ 
\lambda_6 = -\frac{13}{9} - \frac{1}{\sqrt{2}}\frac{l}{\gamma_c}i ,
\label{eq:eigenvallargel2}
\end{equation}
while in the limit of  $l \ll \gamma_c$, 
\begin{equation}
\lambda_1 = -1 ,
\ \ \ 
\lambda_2 = -2 + \frac{1}{3}\frac{l^2}{\gamma_c^2} ,
\ \ \ 
\lambda_3 = -1 - \frac{1}{3}\frac{l^2}{\gamma_c^2} ,
\label{eq:eigenvalsmalll1}
\end{equation}
\begin{equation}
\lambda_4 = -1 - \frac{1}{2}\frac{l^2}{\gamma_c^2} ,
\ \ \ 
\lambda_5 = -2 - \frac{1}{3\sqrt{2}}\frac{l}{\gamma_c} ,
\ \ \ 
\lambda_6 = -2 + \frac{1}{3\sqrt{2}}\frac{l}{\gamma_c} .
\label{eq:eigenvalsmalll2}
\end{equation}

We have calculated the corresponding eigenvectors numerically as shown in
Figure 3 (for $\gamma_c=500$). Each row in the figure contains two panels
which show the real and imaginary parts of one of the six components of the
eigenvectors as functions of $l$. Different line types correspond to the
six different eigenvectors. The complex eigenvectors are all scaled to have
a unit magnitude.  Since each eigenvector corresponds to a mode, the
relative values of the six components of the eigenvector measure the
physical significance of perturbations in different physical parameters for
that mode. For example, the mode corresponding to the eigenvalue
$\lambda_1=-1$ has the temporal behavior of $t^{-1}$; the eigenvector for
this mode is (-0.275, 0.275, -0.824, 0, 0.412, 0), implying that this mode
does not involve $V_T$ and $V_{T3}$ perturbations. Also note that this mode
does not depend on $l$ while all other modes change with $l$.

Equations (\ref{eq:pert_DeltaR_a_wind})--(\ref{eq:pert_F_a_wind}) can also
be solved numerically. We normalize all initial values of the perturbation
variables to unity. The temporal interval of the calculation is from $t=1$
to 30, and $\gamma_c$ is chosen to be 500. The time $t=1$ marks the
beginning of the double shock system.  After the initial explosion, the
fireball expands and accelerates to a relativistic speed. A cold shell is
formed after the thermal energy of the fireball is converted to a bulk
kinetic energy.  The forward/reverse shock system is formed at radius
$R_{t=1}$ where the circumburst medium starts to decelerate significantly
the relativistic shell.  This radius can be approximated as the radius at
which the mass of the circumburst medium swept by the shell is comparable
to $M_0/\eta$, where $M_0$ is the mass of the initial baryonic load of the
fireball, and $\eta$ is the initial thermal Lorentz factor of the
fireball. If the fireball has a total equivalent isotropic energy $E$, we
have $M_0 = E/\eta c^2$. Thus $t=1$ corresponds to a time $T_{t=1} \approx
(1+z)R_{t=1}/2\gamma_c^2 c $ after the GRB trigger in the observer frame,
where $z$ is the cosmological redshift of the source.  Figure 4 shows our
results. The six panels show the time evolution of $\Delta R$, $d_t \Delta
R$, $\delta$, $V_T$, $\delta_3$ and $V_{T3}$.  We show results for four
different $l$ values, namely $l=5$, $l=50$, $l=500$ and $l=5\times
10^3$. These plots indicate that all perturbation variables decay quickly
with time. For small values of $l$ (e.g., $l=5$ and $l=50$) there are no
oscillations. For large values of $l$ (e.g., $l=5\times 10^3$) the
oscillations exist but damp away quickly. These results are consistent with
our analytical derivations. For $l=500$, the oscillations start to appear
although they are not apparent in the plot because of their low
frequency. For $l \gg \gamma_c$, we can calculate the frequencies of the
oscillations using the eigenvalues listed in equations
(\ref{eq:eigenvallargel1}) and (\ref{eq:eigenvallargel2}). The oscillations
have the form $\exp (i\frac{1}{\sqrt{3}}\frac{l}{\gamma_c}\ln t)$ or $\exp
(i\frac{1}{\sqrt{2}}\frac{l}{\gamma_c}\ln t)$, so that the oscillation
period increases with time.  For $l=10\gamma_c$, the shortest period is
$\sim 2.4$, corresponding to $2.4T_{t=1}$ in the observer frame. Hence, the
oscillations could produce fluctuations in the observed flux on time scales
as short as a few times the starting time of the double shock system
$T_{t=1}$.

In the ISM case, $\gamma_c$ is not constant and so we can not write
equations (\ref{eq:pert_DeltaR_a})--(\ref{eq:pert_F_a}) in a matrix form
that admits an analytic solution. Instead, we had to solve these equations
numerically. In order to test the validity of the perturbation equations
and the numerical code, we compared the numerical results for a spherical
perturbation with $l=0$ to the analytic solution derived by perturbing
directly the radial equations of motion, and found an excellent agreement
between the two calculations.  Similarly to the wind case, we normalized
all initial values of the perturbation variables to unity, and chose an
initial $\gamma_c= 500$. Our numerical results are shown in Figure 5 and
resemble qualitatively the wind case. Overall, the perturbations decay
rapidly with time and oscillations appear only for large values of $l$.
Similarly to the wind case, for $l=10\gamma_c$ the shortest period of the
oscillations is $\sim 2$, corresponding to $\sim 2 T_{t=1}$ in the observer
frame. If the equivalent isotropic energy of the fireball is $E=10^{52}$
ergs, the initial thermal Lorentz factor of the fireball is $\eta=10^3$,
the number density of the ISM is $n= 1\ {\rm cm}^{-3}$, then we have
$R_{t=1} = (3E/4\pi n m_p \eta^2 c^2)^{1/3} \sim 1.2\times 10^{16}$
cm. This corresponds to $T_{t=1} = (1+z)R_{t=1}/2\gamma_c^2 c \sim
0.8(1+z)$ sec.  Thus, the time scales of the fluctuations in the observed
flux could be as short as $\sim 2(1+z)$ sec.

\section{Discussion}

We have solved the perturbation equations describing the double
(forward/reverse) shock system which forms during the impact of a highly
relativistic fireball on a surrounding medium. For both a uniform and a
wind ($1/r^2$) density profile of the ambient medium, we have found the
shock system to be stable to global perturbations.  We therefore do not
expect the shock to fragment. Our results are limited to relativistic
reverse shocks, and appear to differ qualitatively from previous results in
the non-relativistic regime (Vishniac 1983).

Our results apply also to collimated outflows as long as the double shock
system is formed at a time when the Lorentz factor of the outflow is larger
than the collimation angle.

We derived the frequencies of the normal modes which could modulate the
short-term variability at the early phase of GRB afterglows.  The results
imply that perturbations in the double shock system could produce
fluctuations in the observed flux on time scales as short as a few seconds
for $\gamma_c\sim 500$ in the ISM case. 
These short term fluctuations could be supplemented by variability on much
longer time scales due to density inhomogeneities in the ISM; such
inhomogeneities can lead to variability on time scales of tens of minutes
in the optical band and days in the radio (Wang \& Loeb 2000).

\acknowledgements

This work was supported in part by grants from the Israel-US BSF
(BSF-9800343), NSF (AST-9900877; AST-0071019), and NASA (NAG5-7039;
NAG5-7768).

\vfil
\eject

\centerline{\bf APPENDIX}
\begin{appendix}

Here we provide full details for the derivation of equations
(\ref{eq:dt_sig_5})--(\ref{eq:dtVT5}) in \S 2.  We start by listing the
equations of motion for a relativistic fluid in spherical coordinates. The
continuity equation reads
\begin{equation}
\frac{\partial}{\partial t}(\gamma \rho) + \frac{1}{r^2}
\frac{\partial}{\partial r}(r^2\gamma \rho u_r) + \frac{1}{r \sin
\theta}\frac{\partial}{\partial \theta} (\sin \theta \gamma \rho
u_{\theta}) + \frac{1}{r \sin \theta}\frac{\partial}{\partial \phi}(\gamma
\rho u_{\phi}) = 0 ,
\label{eq:cont}
\end{equation}
and the three components of the momentum equation are
\begin{equation}
\frac{\gamma^2}{c^2}(e+p)[\frac{\partial u_r}{\partial t} + 
u_r \frac{\partial u_r}{\partial r} + 
u_{\theta} \frac{1}{r}\frac{\partial u_r}{\partial \theta} +
u_{\phi} \frac{1}{r \sin \theta}\frac{\partial u_r}{\partial \phi} - 
\frac{1}{r}(u_{\theta}^2 + u_{\phi}^2)] + \frac{\partial p}{\partial r} 
+ \frac{u_r}{c^2}\frac{\partial p}{\partial t} = 0 ,
\label{eq:mome1}
\end{equation}
\begin{equation}
\frac{\gamma^2}{c^2}(e+p)[\frac{\partial u_{\theta}}{\partial t} + u_r
\frac{\partial u_{\theta}}{\partial r} + u_{\theta}
\frac{1}{r}\frac{\partial u_{\theta}}{\partial \theta} + u_{\phi}
\frac{1}{r \sin \theta}\frac{\partial u_{\theta}}{\partial \phi} +
\frac{1}{r}(u_r u_{\theta} + \cot \theta u_{\phi}^2)] + \frac{1}{r}
\frac{\partial p}{\partial \theta} + \frac{u_{\theta}}{c^2}\frac{\partial
p}{\partial t} = 0 ,
\label{eq:mome2}
\end{equation}
\begin{equation}
\frac{\gamma^2}{c^2}(e+p)[\frac{\partial u_{\phi}}{\partial t} + u_r
\frac{\partial u_{\phi}}{\partial r} + u_{\theta} \frac{1}{r}\frac{\partial
u_{\phi}}{\partial \theta} + u_{\phi} \frac{1}{r \sin \theta}\frac{\partial
u_{\phi}}{\partial \phi} + \frac{1}{r}(u_r u_{\phi} + \cot \theta
u_{\theta} u_{\phi})] + \frac{1}{r \sin \theta} \frac{\partial p}{\partial
\phi} + \frac{u_{\phi}}{c^2}\frac{\partial p}{\partial t} = 0 ,
\label{eq:mome3}
\end{equation}
where $\rho$, $e$, $p$ and $\gamma$ are the fluid density, energy density,
pressure and Lorentz factor respectively, $u_r$, $u_{\theta}$ and
$u_{\phi}$ are the three components of the fluid velocity.

For the forward/reverse shock system under consideration (see Figure
1), we define the following shell--averaged variables for region 2:
\begin{equation}
\sigma(\theta, \phi)=R_0^{-2}\int_{R_0}^{R_1}\gamma \rho r^2 dr , 
\label{eq:sigma_appendix}
\end{equation}
\begin{equation}
V_r(\theta, \phi)=(\sigma R_0^2)^{-1}\int_{R_0}^{R_1}\gamma \rho u_r r^2 dr ,
\label{eq:Vr_appendix}
\end{equation}
\begin{equation}
{\bf V}_T(\theta, \phi)=(\sigma R_0^2)^{-1}\int_{R_0}^{R_1} \gamma \rho
{\bf u}_T r^2 dr ,
\label{eq:VT_appendix}
\end{equation}
where ${\bf u}_T$ is the tangential velocity vector.

Since the shocked material is relativistic, we adopt the relativistic
equation of state, $p=e/3$, in region 2.  Equation (\ref{eq:cont}) yields
the evolution of the surface density
\begin{eqnarray}
\partial_t \sigma & = & -2\frac{\dot{R}_0}{R_0}\sigma +
\left(\frac{R_1}{R_0}\right)^2\gamma(R_1)\rho(R_1)\dot{R}_1 -
\gamma(R_0)\rho(R_0)\dot{R}_0 +
R_0^{-2}\int_{R_0}^{R_1}\frac{\partial}{\partial t}(\gamma \rho) r^2 dr
\nonumber\\ & = & -2\frac{\dot{R}_0}{R_0}\sigma +
\left(\frac{R_1}{R_0}\right)^2\gamma(R_1)\rho(R_1)[\dot{R}_1 - u_r(R_1)] +
\gamma(R_0)\rho(R_0)[u_r(R_0) - \dot{R}_0] \nonumber\\ & & -
R_0^{-2}\int_{R_0}^{R_1} [{\bf \nabla} \cdot (\gamma \rho {\bf u}_T)] r^2
dr .
\label{eq:dt_sig}
\end{eqnarray}
Since $r = R_0$ defines the contact discontinuity between the shocked shell
and the shocked circumburst medium and there is no mass flow across the
contact discontinuity, we get $u_r(R_0) = \dot{R}_0$. Because the forward
shock is a strong relativistic shock, we have the following shock jump
conditions at shock 1:
\begin{equation}
\gamma^2(R_1) = \gamma_{s1}^2/2 ,
\label{eq:shock1a}
\end{equation}
\begin{equation}
\rho(R_1)/\rho_1 = 4\gamma(R_1) ,
\label{eq:shock1b}
\end{equation}
where $\gamma(R_1)$ and $\rho(R_1)$ are the Lorentz factor and density of
the fluid just behind the shock front, $\gamma_{s1}$ is the Lorentz factor
of the shock front and $\rho_1$ is the density of the unshocked circumburst
medium just in front of the shock front. In the highly relativistic regime,
\begin{equation}
\dot{R}_1 \approx c\left( 1-\frac{1}{2\gamma_{s1}^2}\right) ,
\label{eq:R1dot}
\end{equation}
\begin{equation}
u_r(R_1) \approx c\left(1-\frac{1}{2\gamma^2(R_1)}\right) .
\label{eq:urR1}
\end{equation}
 From equations (\ref{eq:shock1a})--(\ref{eq:urR1}), we obtain 
\begin{equation}
\gamma(R_1)\rho(R_1)[\dot{R}_1 - u_r(R_1)] \approx \rho_1 c .
\label{eq:bound1}
\end{equation}
Thus, equation (\ref{eq:dt_sig}) can be rewritten as 
\begin{equation}
\partial_t \sigma = -2\frac{\dot{R}_0}{R_0}\sigma +
\left(\frac{R_1}{R_0}\right)^2\rho_1c - R_0^{-2}\int_{R_0}^{R_1} [{\bf
\nabla} \cdot (\gamma \rho {\bf u}_T)] r^2 dr .
\label{eq:dt_sig_2}
\end{equation}
To linear order, the last integration term in the above equation 
can be approximated by 
\begin{equation}
- R_0^{-2}\int_{R_0}^{R_1} [{\bf \nabla} \cdot (\gamma \rho {\bf u}_T)] r^2
dr = - \sigma {\bf \nabla}_T \cdot {\bf V}_T + R_0^{-2}\int_{R_0}^{R_1}
[{\bf \nabla} \cdot (\gamma \rho {\bf u}_T)] \left(\frac{r}{R_0}-1\right)
r^2 dr ,
\label{eq:int1}
\end{equation}
where the operator ${\bf \nabla}_T \equiv (\hat{\theta}/R_0)(\partial
/\partial \theta) + (\hat{\phi}/R_0)(\partial/\partial \phi)$ 
acts as follows on a scalar $\Psi$ and a vector ${\bf f}$: 
\begin{equation}
{\bf \nabla}_T\Psi = \frac{1}{R_0}\frac{\partial \Psi}
{\partial \theta} \hat{\theta}
+\frac{1}{R_0\sin \theta}\frac{\partial \Psi}
{\partial \phi} \hat{\phi} ,
\label{eq:deltaTgradient}
\end{equation}
\begin{equation}
{\bf \nabla}_T \cdot {\bf f} = 
\frac{1}{R_0 \sin \theta}\frac{\partial}{\partial \theta} 
(\sin \theta f_{\theta})
+\frac{1}{R_0\sin \theta}\frac{\partial f_{\phi}}
{\partial \phi} .
\label{eq:deltaTdivergence}
\end{equation}
Note that the second term on the right-hand side of equation
(\ref{eq:int1}) is of higher order in $(R_1-R_0)/R_0$ than the preceding
term, and hence can be ignored in the thin shell approximation. Thus
equation (\ref{eq:dt_sig_2}) can be rewritten as
\begin{equation}
\partial_t \sigma = -2\frac{\dot{R}_0}{R_0}\sigma 
+ \left(\frac{R_1}{R_0}\right)^2\rho_1c 
- \sigma {\bf \nabla}_T \cdot {\bf V}_T .
\label{eq:dt_sig_3}
\end{equation}

Similarly to the above derivation, we obtain for the bulk radial velocity
\begin{eqnarray}
\partial_t V_r(\theta, \phi) & = & -\frac{\partial_t \sigma}{\sigma}V_r 
- 2 \frac{\dot{R}_0}{R_0}V_r 
+ \sigma^{-1}\left(\frac{R_1}{R_0}\right)^2\gamma(R_1)\rho(R_1)u_r(R_1)
[\dot{R}_1 - u_r(R_1)]
\nonumber\\ 
 & & + \sigma^{-1}\gamma(R_0)\rho(R_0)u_r(R_0)[u_r(R_0) - \dot{R}_0] 
- (\sigma R_0^2)^{-1}\int_{R_0}^{R_1} 
[{\bf \nabla} \cdot (\gamma \rho {\bf u}_T)] u_r r^2 dr 
\nonumber\\
 & & - (\sigma R_0^2)^{-1}\int_{R_0}^{R_1}\frac{\rho c^2}{4 \gamma p} 
\left( \frac{\partial p}{\partial r} 
+ \frac{u_r}{c^2}\frac{\partial p}{\partial t} \right) r^2 dr 
- (\sigma R_0^2)^{-1}\int_{R_0}^{R_1}\gamma \rho 
({\bf u}_T \cdot {\bf \nabla} u_r) r^2 dr 
\nonumber\\ 
 & & + (\sigma R_0^2)^{-1}\int_{R_0}^{R_1}\gamma \rho {\bf u}_T^2 r dr .
\label{eq:dtVr}
\end{eqnarray}
The last two terms on the right-hand-side of the above equation are
nonlinear. By substituting equations (\ref{eq:bound1}) and
(\ref{eq:dt_sig_2}) into the above equation, and keeping terms
to the linear order we get
\begin{eqnarray}
\partial_t V_r(\theta, \phi) & = & 
\sigma^{-1}\left(\frac{R_1}{R_0}\right)^2 \rho_1 c [u_r(R_1) - V_r] 
\nonumber\\
 & & 
+ \left[ (\sigma R_0^2)^{-1}V_r\int_{R_0}^{R_1}
 [{\bf \nabla} \cdot (\gamma \rho {\bf u}_T)] r^2 dr
- (\sigma R_0^2)^{-1}\int_{R_0}^{R_1}u_r 
 [{\bf \nabla} \cdot (\gamma \rho {\bf u}_T)] r^2 dr \right] 
\nonumber\\
 & & 
- (\sigma R_0^2)^{-1}\int_{R_0}^{R_1}\frac{\rho c^2}{4 \gamma p} 
\left( \frac{\partial p}{\partial r} 
+ \frac{u_r}{c^2}\frac{\partial p}{\partial t} \right) r^2 dr .
\label{eq:dtVr2}
\end{eqnarray}
In order to evaluate the integral in the last term of the above equation,
we need the relation between $p$ and $\rho$ inside region 2. Since entropy 
is conserved in this region,
\begin{equation}
\frac{d}{dt}\left(\frac{p}{\rho^{4/3}}\right)=0 ,
\label{eq:ener_new}
\end{equation}
implying that $p/\rho^{4/3}$ remains constant for a given fluid
particle. Hence, a fluid layer which is at a distance $x(t)$ from the
contact discontinuity ($r=R_0$) inside region 2, maintains a constant
$p/\rho^{4/3}$ over time and its value is decided by the Lorentz factor of
shock 1 at the time when this layer first crossed shock 1.  However, at a
particular time, different layers across region 2 have different values of
$p/\rho^{4/3}$. Assuming that region 2 is decelerating with $\gamma \propto
r^{-1/2}$ (for the uniform ISM case), we get
\begin{equation}
\frac{p(x)}{\rho^{4/3}(x)}=\frac{c^2}{3\cdot 4^{1/3}\rho_1^{1/3}}
\left[\frac{\gamma_a R_a^{1/2}}{(8\gamma_a^2 R_a x +
R_a^2)^{1/4}}\right]^{2/3} ,
\label{eq:p_rho}
\end{equation}
where $\gamma_a$ and $R_a$ are the Lorentz factor and radius of region 2 at
the initial time.  Apparently, the dependence of $p(x)/ \rho^{4/3}(x)$ on
$x$ is very weak, and so within the context of the thin shell approximation
we simply assume that $p/ \rho^{4/3}$ is constant across region 2 at any
given time.  This assumption is indeed satisfied in the numerical
simulations performed by Kobayashi, Piran, \& Sari (1999).  In equation
(\ref{eq:p_rho}), the term $8\gamma_a^2 R_a x$ can be at most comparable to
$R_a^2$ (this happens in the very last stage of the evolution when the
reverse shock crosses the shell), and so the error introduced by our
approximation is small.  For the wind case, $\gamma \approx {\rm const}$,
and we can also treat $p/ \rho^{4/3}$ as a constant across region 2.

We may now calculate the integral
\begin{eqnarray} 
\int_{R_0}^{R_1}\frac{\rho c^2}{4 \gamma p} \frac{\partial p}{\partial r}
r^2 dr & = & \int_{R_0}^{R_1}\frac{c^2}{3\gamma}\frac{\partial
\rho}{\partial r} r^2 dr =
\frac{c^2}{3\gamma(R_0)}\int_{R_0}^{R_1}\frac{\partial \rho}{\partial r}
r^2 dr \nonumber\\ & \approx &
\frac{c^2}{3\gamma(R_0)}\left[\rho(R_1)R_1^2-\rho(R_0)R_0^2 -
\int_{R_0}^{R_1}2r\rho dr \right] .
\label{eq:int_p_r}
\end{eqnarray}
In the thin shell approximation, all radial velocities are dominated by the
overall radial motion of the shock layer. Hence, $\gamma$ was treated as a
constant and was taken out of the integral. For the uniform ISM case, the
density difference between the two edges of region 2 is not small; hence,
the last term inside the square brackets is of order $(R_1-R_0)/R_0$ times
the difference between the previous two terms and so it can be
neglected. For the wind case, there is no pressure gradient across region
2, and so the above integral vanishes. However, even in this case we can
ignore the last term and keep only the first two terms, because later on we
will replace $R_1$ with $R_0$, and so the difference between the first two
terms will vanish.

Another relevant integral is
\begin{eqnarray} 
\int_{R_0}^{R_1}\frac{\rho c^2}{4 \gamma p} \frac{u_r}{c^2} \frac{\partial
p}{\partial t} r^2 dr & = & \int_{R_0}^{R_1}\frac{c^2}{3\gamma}
\frac{u_r}{c^2} \frac{\partial \rho}{\partial t} r^2 dr =
\frac{V_r}{3\gamma(R_0)}\int_{R_0}^{R_1}\frac{\partial \rho}{\partial t}
r^2 dr \nonumber\\ & = & \frac{V_r}{3\gamma(R_0)}\left[
\frac{\partial}{\partial t}\int_{R_0}^{R_1}\rho r^2 dr
-\dot{R}_1\rho(R_1)R_1^2 + \dot{R}_0\rho(R_0)R_0^2 \right] \nonumber\\ & =
& \frac{V_r}{3\gamma(R_0)}\left[ \frac{\sigma}{\gamma(R_0)}2R_0\dot{R}_0 +
\frac{R_0^2}{\gamma(R_0)}\partial_t\sigma
-\frac{R_0^2\sigma}{\gamma^2(R_0)}\partial_t\gamma(R_0) \right.
\nonumber\\ & & \left. -\dot{R}_1\rho(R_1)R_1^2 + \dot{R}_0\rho(R_0)R_0^2
\right] .
\label{eq:int_p_t}
\end{eqnarray}
In the above derivation we pulled $u_r$ out of the integration assuming
that it equals $V_r$, as appropriate in the thin shell approximation.  By
substituting equation (\ref{eq:dt_sig_3}) into equation (\ref{eq:int_p_t})
and making use of the following two relations
\begin{equation}
c^2-V_r\dot{R}_0 \approx \frac{c^2}{\gamma^2(R_0)} ,
\label{eq:taylor1}
\end{equation}
\begin{equation}
c^2-V_r\dot{R}_1 \approx \frac{3c^2}{4\gamma^2(R_0)} , 
\label{eq:taylor2}
\end{equation}
we get
\begin{eqnarray}
\int_{R_0}^{R_1}\frac{\rho c^2}{4 \gamma p} \left(\frac{\partial
p}{\partial r} + \frac{u_r}{c^2}\frac{\partial p}{\partial t} \right)r^2 dr
& = &\frac{4R_0^2}{3\gamma^2(R_0)}\rho_1c^2 -
\frac{R_0^2}{3\gamma^3(R_0)}\rho(R_0)c^2 \nonumber\\ & & -\sigma
\frac{R_0^2}{3\gamma^2(R_0)}V_r{\bf \nabla}_T \cdot {\bf V}_T -\sigma
\frac{R_0^2}{3\gamma^3(R_0)}V_r \partial_t\gamma(R_0) .
\label{eq:int_com}
\end{eqnarray}
Thus, equation~(\ref{eq:dtVr2}) is now changed to
\begin{eqnarray}
\partial_t V_r & = & \sigma^{-1}\left(\frac{R_1}{R_0}\right)^2 \rho_1 c
[u_r(R_1) - V_r] \nonumber\\ & & + \left[ (\sigma
R_0^2)^{-1}V_r\int_{R_0}^{R_1} [{\bf \nabla} \cdot (\gamma \rho {\bf u}_T)]
r^2 dr - (\sigma R_0^2)^{-1}\int_{R_0}^{R_1}u_r [{\bf \nabla} \cdot (\gamma
\rho {\bf u}_T)] r^2 dr \right] \nonumber\\ & &
-\sigma^{-1}\frac{4}{3\gamma^2(R_0)}\rho_1c^2
+\sigma^{-1}\frac{1}{3\gamma^3(R_0)}\rho(R_0)c^2
+\frac{1}{3\gamma^2(R_0)}V_r{\bf \nabla}_T \cdot {\bf V}_T \nonumber\\ & &
+\frac{1}{3\gamma^3(R_0)}V_r \partial_t\gamma(R_0) .
\label{eq:dtVr3}
\end{eqnarray}
In order to close the final equations, we can only have one free variable
for the radial velocities. We use the approximation that $u_r$ is constant
across region 2 with $V_r = u_r(R_1)$, as appropriate under the thin shell
approximation. Hence, the first two terms in equation ~(\ref{eq:dtVr3})
both vanish, and we end up with the following equation,
\begin{equation}
\partial_t V_r = 
-\sigma^{-1}\frac{4}{3\gamma^2(R_0)}\rho_1c^2 
+\sigma^{-1}\frac{1}{3\gamma^3(R_0)}\rho(R_0)c^2
+\frac{1}{3\gamma^2(R_0)}V_r{\bf \nabla}_T \cdot {\bf V}_T
+\frac{1}{3\gamma^3(R_0)}V_r \partial_t\gamma(R_0) .
\label{eq:dtVr4}
\end{equation}
Since 
\begin{equation}
\frac{\partial_t\gamma(R_0)}{\gamma^3(R_0)}=\frac{V_r}{c^2}\partial_t V_r ,
\label{eq:gammaVr}
\end{equation}
equation~(\ref{eq:dtVr4}) can be rewritten as
\begin{equation}
\partial_t\gamma(R_0)
= -2\sigma^{-1}\gamma(R_0)\rho_1c
+\frac{1}{2}\sigma^{-1}\rho(R_0)c
+\frac{\gamma(R_0)}{2c}V_r{\bf \nabla}_T \cdot {\bf V}_T .
\label{eq:dtVr5}
\end{equation}
Because $p/ \rho^{4/3}$ is a constant across region 2, 
we obtain the  following relation
\begin{equation}
\rho(R_0) = 3^{3/4}\cdot 2^{1/2} \frac{p^{3/4}(R_0)\rho_1^{1/4}}
{\gamma^{1/2}(R_0)c^{3/2}} .
\label{eq:rhor0}
\end{equation}
Using this result, we can rewrite equation~(\ref{eq:dtVr5}) as
\begin{equation}
\partial_t\gamma(R_0)
= -2\sigma^{-1}\gamma(R_0)\rho_1c
+\frac{3^{3/4}}{2^{1/2}}\sigma^{-1}\frac{p^{3/4}(R_0)\rho_1^{1/4}}
{\gamma^{1/2}(R_0)c^{1/2}}
+\frac{\gamma(R_0)}{2c}V_r{\bf \nabla}_T \cdot {\bf V}_T .
\label{eq:dtVr6}
\end{equation}

For the tangential velocity we have 
\begin{eqnarray}
\partial_t {\bf V}_T(\theta, \phi) & = & 
-\frac{\partial_t \sigma}{\sigma} {\bf V}_T
- 2 \frac{\dot{R}_0}{R_0} {\bf V}_T
+ \sigma^{-1}\left(\frac{R_1}{R_0}\right)^2\gamma(R_1)\rho(R_1)
{\bf u}_T(R_1) [\dot{R}_1 - u_r(R_1)]
\nonumber\\ 
 & & + \sigma^{-1}\gamma(R_0)\rho(R_0){\bf u}_T(R_0)[u_r(R_0) - \dot{R}_0] 
- (\sigma R_0^2)^{-1}\int_{R_0}^{R_1} 
\gamma \rho \frac{u_r {\bf u}_T}{r}r^2dr 
\nonumber\\
 & & - (\sigma R_0^2)^{-1}\int_{R_0}^{R_1}\frac{\rho c^2}{4 \gamma p} 
\left({\bf \nabla}_T p + \frac{{\bf u}_T}{c^2}
\frac{\partial p}{\partial t}\right) r^2 dr 
\nonumber\\
 & & - (\sigma R_0^2)^{-1}\int_{R_0}^{R_1}\gamma \rho 
({\bf u}_T \cdot {\bf \nabla} {\bf u}_T) r^2 dr
- (\sigma R_0^2)^{-1}\int_{R_0}^{R_1}
{\bf u}_T \nabla \cdot (\gamma \rho {\bf u}_T) r^2 dr .
\label{eq:dtVT}
\end{eqnarray}
Apparently the last two terms in the above equation are nonlinear and can
be ignored. By substituting equations (\ref{eq:bound1}) and
(\ref{eq:dt_sig_3}) into the above equation, we get
\begin{eqnarray}
\partial_t {\bf V}_T & = & \sigma^{-1}\left(\frac{R_1}{R_0}\right)^2
\rho_1c[{\bf u}_T(R_1)-{\bf V}_T] + {\bf V}_T {\bf \nabla}_T \cdot {\bf
V}_T - (\sigma R_0^2)^{-1}\int_{R_0}^{R_1} \gamma \rho \frac{u_r {\bf
u}_T}{r}r^2dr \nonumber\\ & & - (\sigma
R_0^2)^{-1}\int_{R_0}^{R_1}\frac{\rho c^2}{4 \gamma p} \left({\bf \nabla}_T
p + \frac{{\bf u}_T}{c^2} \frac{\partial p}{\partial t}\right) r^2 dr .
\label{eq:dtVT2}
\end{eqnarray}
The second term on the right-hand-side of equation (\ref{eq:dtVT2}) is
nonlinear and can be neglected. In the thin shell approximation, the third
term on the right-hand side of equation (\ref{eq:dtVT2}) can be
approximated as $- V_r {\bf V}_T / R_0$. Using the shock jump conditions at
shock 1 and making use of the fact that the tangential velocities must be
continuous across the shock front, we obtain
\begin{equation}
{\bf u}_T (R_1) = -u_r(R_1) ({\bf \nabla}_T R_1) .
\label{eq:bound2}
\end{equation}
Based on these considerations, equation (\ref{eq:dtVT2}) can be rewritten
as
\begin{eqnarray}
\partial_t {\bf V}_T & = & 
\sigma^{-1}\left(\frac{R_1}{R_0}\right)^2 \rho_1c
[-u_r(R_1) ({\bf \nabla}_T R_1) - {\bf V}_T] 
- \frac{V_r}{R_0}{\bf V}_T 
\nonumber\\
 & & - (\sigma R_0^2)^{-1}\int_{R_0}^{R_1}\frac{\rho c^2}{4 \gamma p} 
\left({\bf \nabla}_T p + \frac{{\bf u}_T}{c^2}
\frac{\partial p}{\partial t}\right) r^2 dr .
\label{eq:dtVT3}
\end{eqnarray}
Next we consider the integration term in the above equation which includes
\begin{eqnarray}
\int_{R_0}^{R_1}\frac{\rho c^2}{4 \gamma p}({\bf \nabla}_T p) r^2 dr & = &
\int_{R_0}^{R_1}\frac{c^2}{3 \gamma}({\bf \nabla}_T \rho) r^2 dr
\nonumber\\ & = & \int_{R_0}^{R_1}\frac{c^2}{3 \gamma^2}[{\bf \nabla}_T
(\gamma \rho)] r^2 dr - \int_{R_0}^{R_1}\frac{\rho c^2}{3 \gamma^2} ({\bf
\nabla}_T \gamma) r^2 dr \nonumber\\ & = & 
\frac{c^2}{3\gamma^2(R_0)}\left[
{\bf \nabla}_T\int_{R_0}^{R_1}\gamma \rho r^2 dr
-\gamma(R_1)\rho(R_1)R_1^2({\bf \nabla}_T R_1) \right. \nonumber\\ & &
\left. +\gamma(R_0)\rho(R_0)R_0^2({\bf \nabla}_T R_0) \right]
-\frac{c^2 {\bf \nabla}_T \gamma(R_0)}{3 \gamma^3(R_0)}
\int_{R_0}^{R_1}\gamma \rho r^2 dr \nonumber\\ & = &
\frac{c^2}{3\gamma^2(R_0)}\left[ R_0^2 {\bf \nabla}_T \sigma
+2\sigma R_0 ({\bf \nabla}_T R_0) -\gamma(R_1)\rho(R_1)R_1^2({\bf \nabla}_T R_1)
\right. \nonumber\\ & &
\left. +\gamma(R_0)\rho(R_0)R_0^2({\bf \nabla}_T R_0) \right]
-\frac{c^2}{3\gamma^3(R_0)} R_0^2\sigma {\bf \nabla}_T \gamma(R_0) .
\label{eq:del_p}
\end{eqnarray}
In the last pair of square brackets of equation~(\ref{eq:del_p}), the
second term is much smaller than the fourth term and so it can be
neglected.  Another integration term is
\begin{eqnarray}
\int_{R_0}^{R_1}\frac{\rho c^2}{4 \gamma p}
\frac{{\bf u}_T}{c^2} \frac{\partial p}{\partial t} r^2 dr 
& = & 
\int_{R_0}^{R_1}\frac{1}{4\gamma^2}(\partial_t\ln p)\gamma\rho{\bf u}_Tr^2 dr
\nonumber\\
& \approx & 
\frac{1}{4\gamma^2(R_0)}[\partial_t\ln p(R_1)]
\int_{R_0}^{R_1}\gamma\rho{\bf u}_Tr^2 dr 
\nonumber\\
& = & 
\frac{\partial_t\gamma(R_0)}{2\gamma^3(R_0)}\sigma R_0^2{\bf V}_T 
+\frac{1}{4\gamma^2(R_0)}\left( \frac{\partial_t \rho_1}{\rho_1}\right)
\sigma R_0^2{\bf V}_T .
\label{eq:u_Tpt1}
\end{eqnarray}
Because $\partial_t\ln p$ does not change much across region 2, 
we took it out of the integration in the above derivation.

By substituting equation (\ref{eq:dtVr5}) into equation (\ref{eq:u_Tpt1}),
we get
\begin{equation}
\int_{R_0}^{R_1}\frac{\rho c^2}{4 \gamma p} \frac{{\bf u}_T}{c^2}
\frac{\partial p}{\partial t} r^2 dr \approx
-\frac{R_0^2}{\gamma^2(R_0)}\rho_1c{\bf V}_T
+\frac{R_0^2}{4\gamma^3(R_0)}\rho(R_0)c{\bf V}_T 
+\frac{1}{4\gamma^2(R_0)}\left( \frac{\partial_t \rho_1}{\rho_1}\right)
\sigma R_0^2{\bf V}_T .
\label{eq:u_Tpt2}
\end{equation}
Now, by substituting equations (\ref{eq:del_p}) and (\ref{eq:u_Tpt2}) into
equation (\ref{eq:dtVT3}), we get
\begin{eqnarray}
\partial_t {\bf V}_T & = & \frac{1}{3}\sigma^{-1}c^2
\left[\rho_1-\frac{\rho(R_0)}{\gamma(R_0)}\right]
{\bf \nabla}_T R_0
-\sigma^{-1}\rho_1c{\bf V}_T
-\frac{V_r}{R_0}{\bf V}_T 
\nonumber \\
& & 
-\sigma^{-1}\frac{c^2}{3\gamma^2(R_0)}{\bf \nabla}_T \sigma 
+ \frac{c^2}{3\gamma^3(R_0)}{\bf \nabla}_T \gamma(R_0) 
-\frac{1}{4\gamma^2(R_0)}\left( \frac{\partial_t \rho_1}{\rho_1}\right)
{\bf V}_T . 
\label{eq:dtVT4}
\end{eqnarray}
In deriving the above equation, we made the assumption that surface
irregularities due to variations in the thickness of the shock layer is of
higher order than irregularities due to the bulk displacement of the shock
layer (regions 2 \& 3 in Figure 1), so ${\bf \nabla}_T R_0 = {\bf \nabla}_T
R_1 = {\bf \nabla}_T R_2$.  This is appropriate under the thin shell
approximation.  Also the last term in the above equation is much smaller
than the third term for the wind case and is equal to zero for the ISM
case, so can be neglected.  If we now substitute equation (\ref{eq:rhor0})
into the above equation, we get equation (\ref{eq:dtVT5}) in \S 2.

The derivation of the perturbation equations of region 3 is very similar to
that of region 2.

\end{appendix}

\newpage
\begin{figure} [htbp]
\centerline{\epsfysize = 4in \epsffile{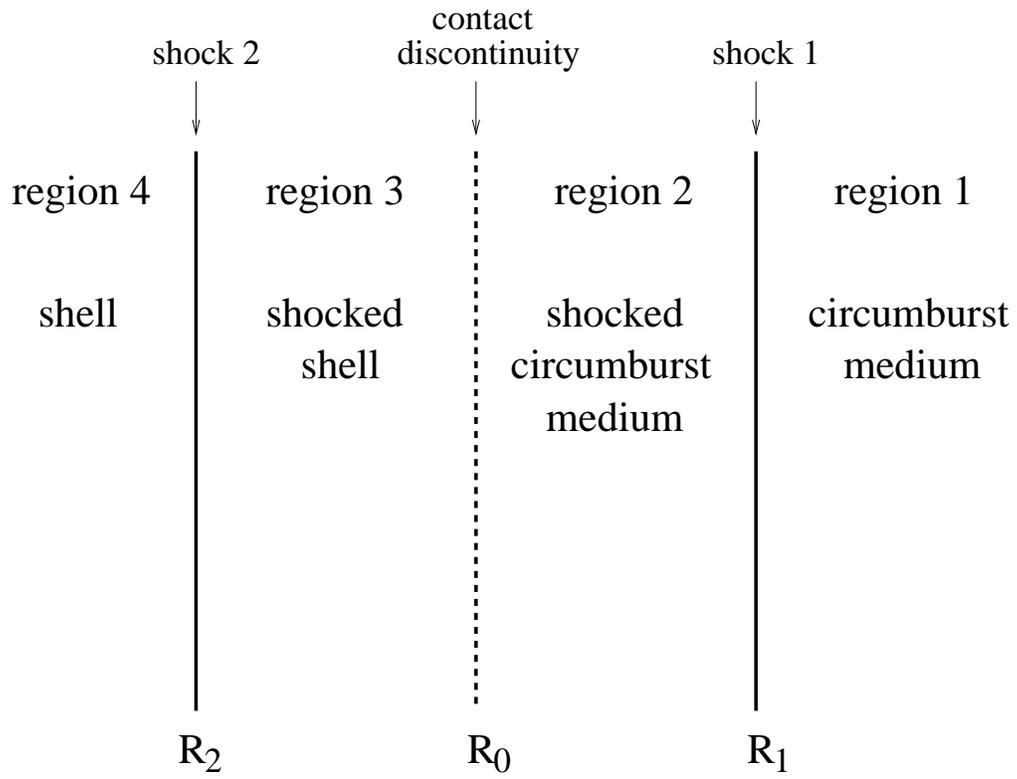}}
\caption{Structure of the forward/reverse shock system.}
\end{figure}

\begin{figure} [htbp]
\includegraphics{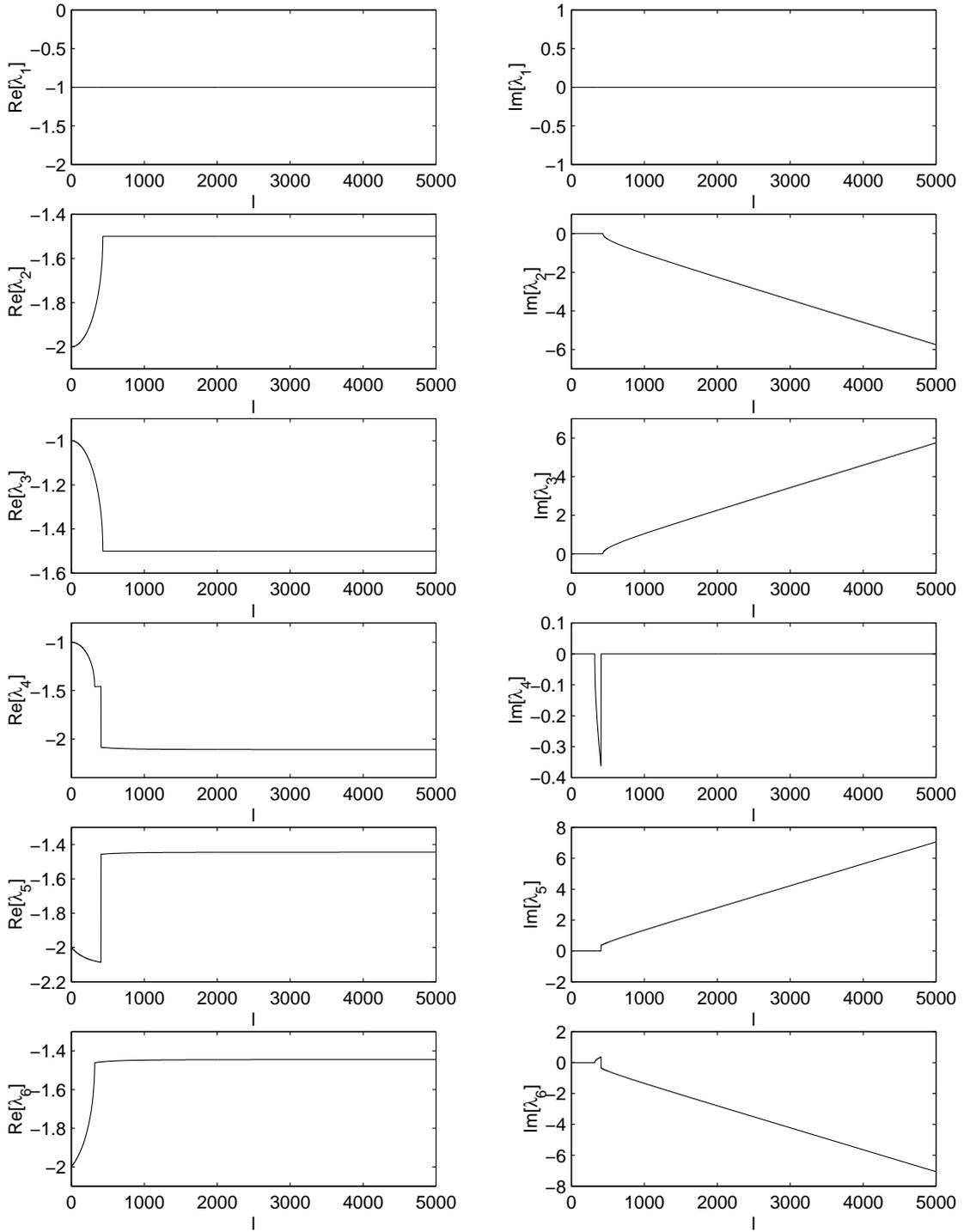}
\vspace{7.0in}
\caption{Real and imaginary parts of the six eigenvalues in the wind case
as functions of $l$ with $\gamma_c=500$.}
\end{figure}

\begin{figure} [htbp]
\includegraphics{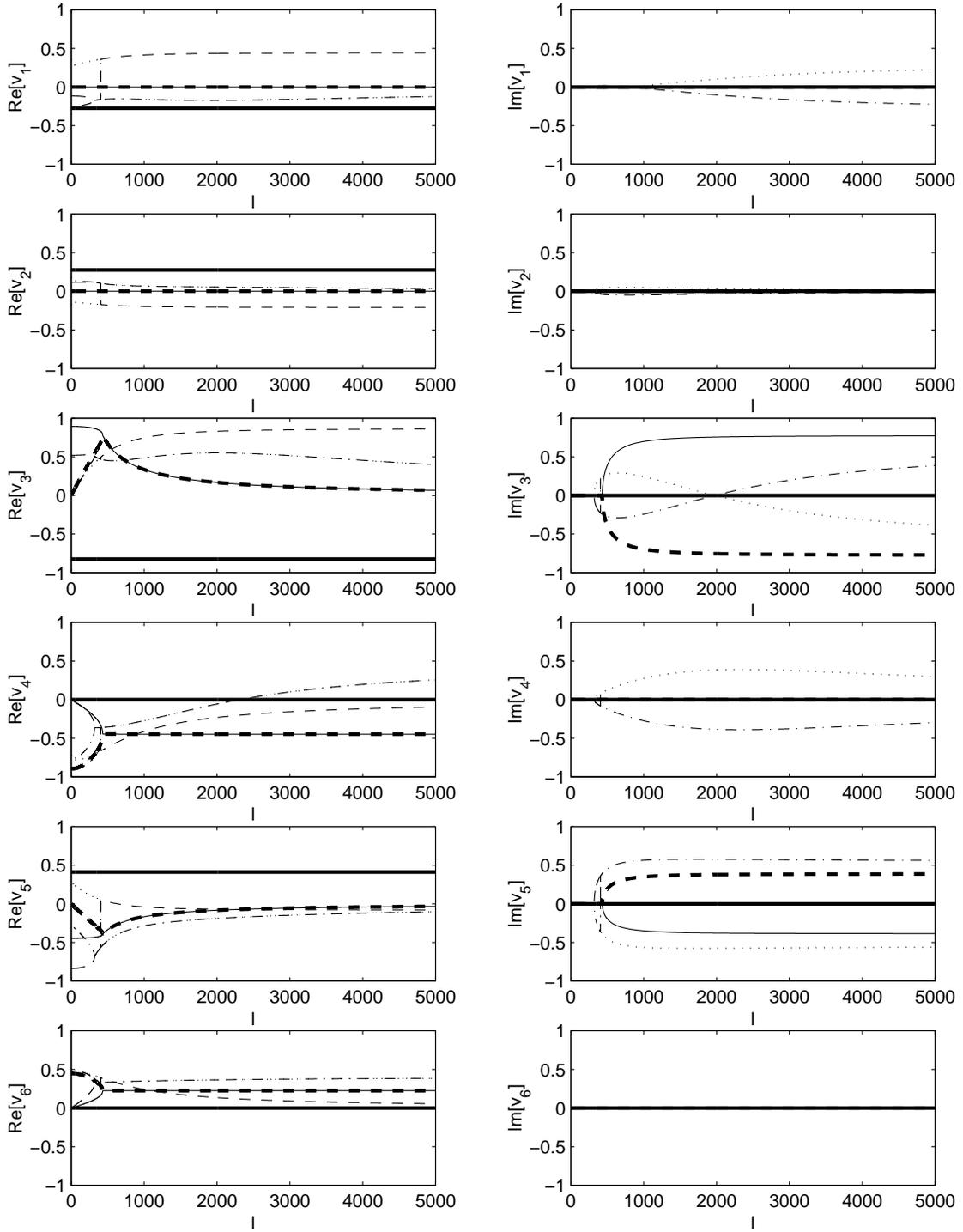}
\vspace{7.0in}
\caption{Real and imaginary parts of the six components of different
eigenvectors (modes) in the wind case with $\gamma_c=500$. 
Each row corresponds to one component of the eigenvector.  Different 
line types correspond to six different eigenvectors: the thick solid 
line refers to the eigenvector of $\lambda_1$, the thick dashed line 
to $\lambda_2$, the thin solid line to $\lambda_3$, the thin dashed 
line to $\lambda_4$, the dotted line to $\lambda_5$ and the dashed-dot 
line refers to $\lambda_6$.}
\end{figure}

\begin{figure} [htbp]
\includegraphics{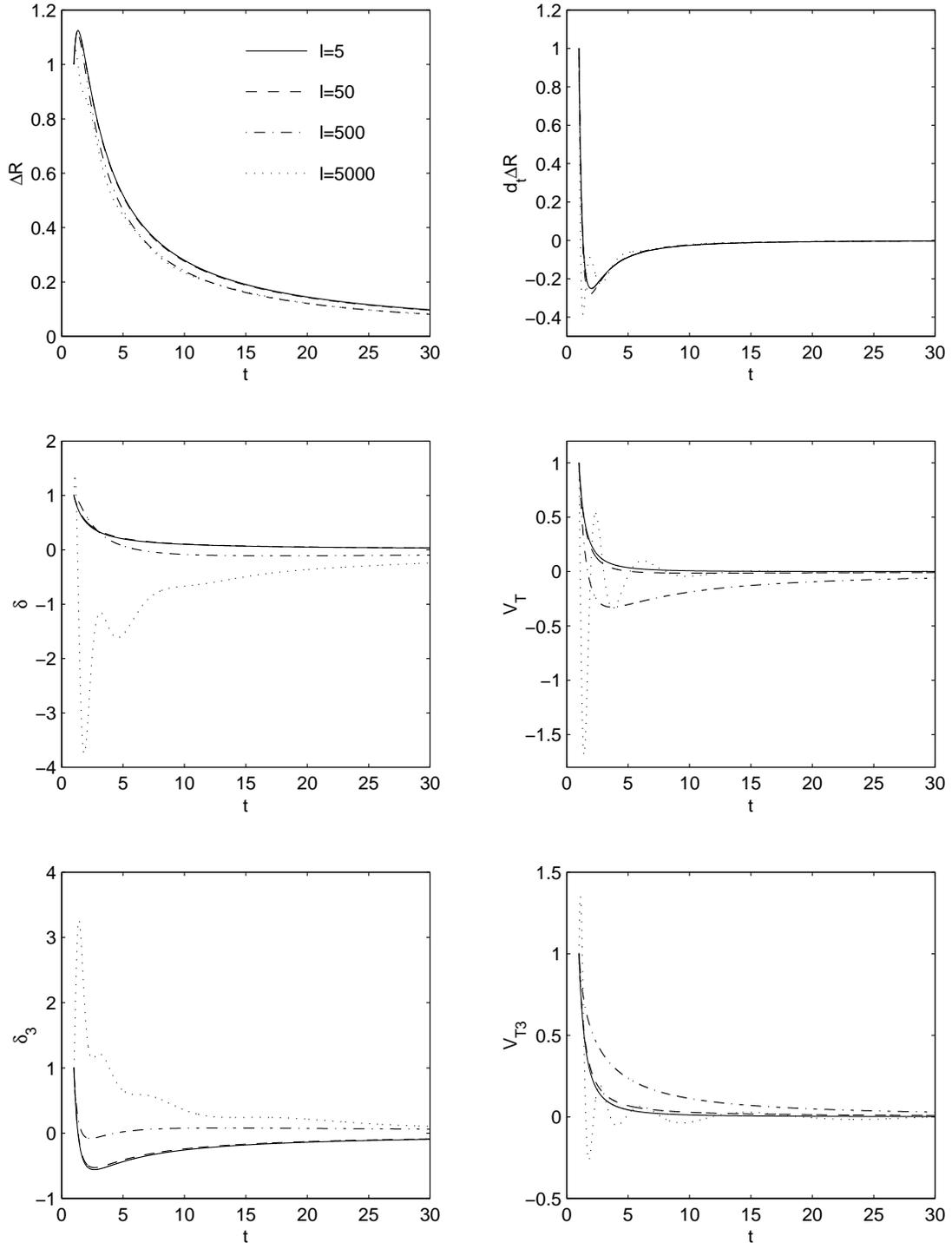}
\vspace{7.0in}
\caption{Evolution of the perturbation variables in the wind case 
with $\gamma_c=500$. Four different line types correspond to $l=5$, 
$l=50$, $l=500$ and $l=5\times 10^3$.}
\end{figure}

\begin{figure} [htbp]
\includegraphics{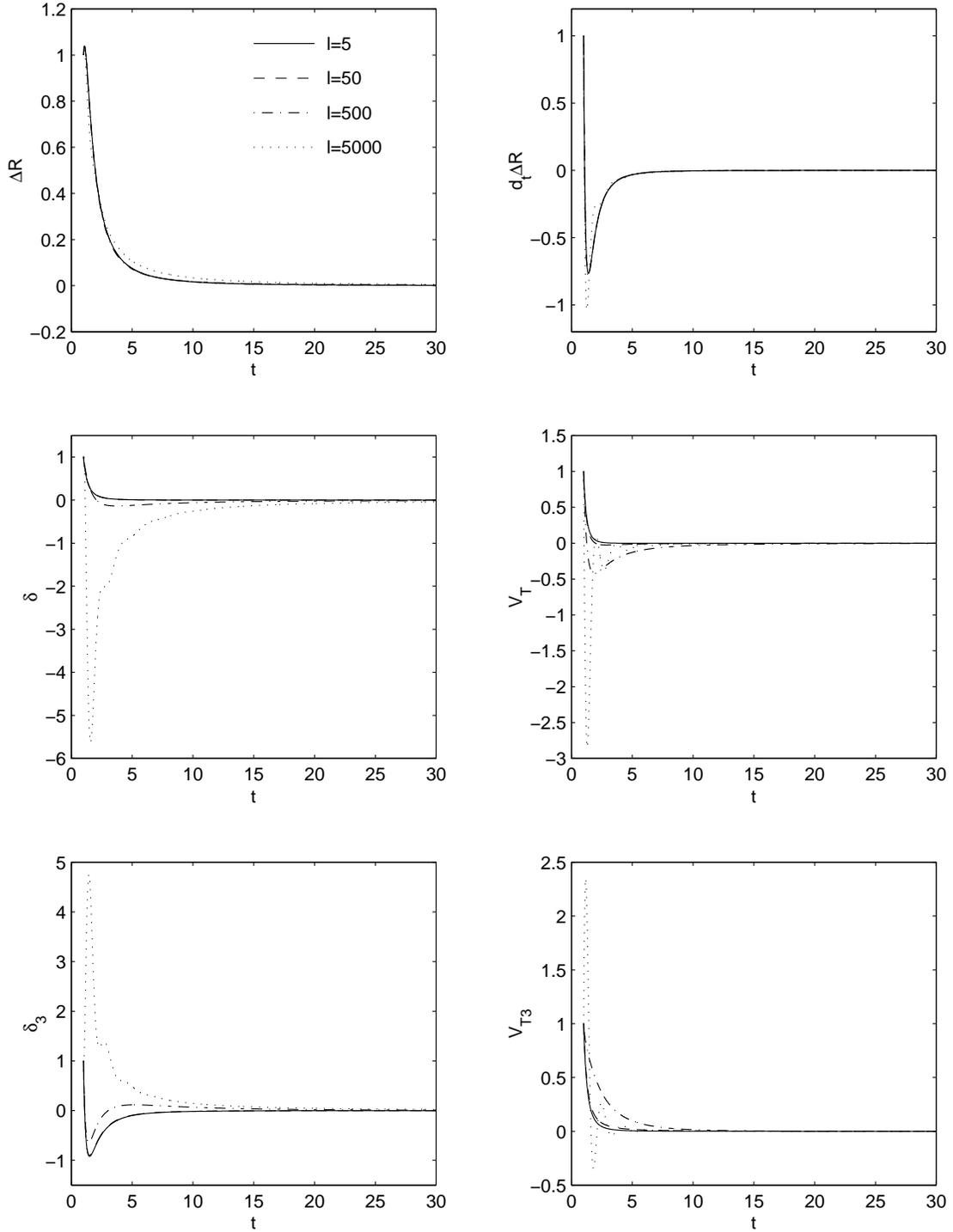}
\vspace{7.0in}
\caption{Evolution of the perturbation variables in the ISM case 
with $\gamma_c=500$.} 
\end{figure}

\end{document}